\pdfoutput=1 %
\documentclass{article}

\usepackage{arxiv}

\usepackage[utf8]{inputenc} %
\usepackage[T1]{fontenc}    %
\usepackage{hyperref}       %
\usepackage{url}            %
\usepackage{booktabs}       %
\usepackage{array}          %
\usepackage{amsmath}        %
\usepackage{amsfonts}       %
\usepackage{nicefrac}       %
\usepackage{microtype}      %
\usepackage{cleveref}       %
\usepackage{graphicx}
\usepackage{listings}      %
\usepackage{upquote}       %
\crefname{lstlisting}{Listing}{Listings}
\Crefname{lstlisting}{Listing}{Listings}
\crefname{appendix}{appendix}{appendices}
\Crefname{appendix}{Appendix}{Appendices}
\crefname{subappendix}{appendix}{appendices}
\Crefname{subappendix}{Appendix}{Appendices}
\usepackage{natbib}
\usepackage{doi}

\title{Repository-Grounded Rubrics for Execution-Free Evaluation of Coding Agents}

\date{November 10, 2026}

\usepackage{authblk}

\author{Matej Gerek\thanks{Corresponding author: \texttt{matejgerek@faros.ai}}}
\author{Danylo Zahorulko}
\author{Samuel Budai}
\author{Chris Rupley}
\author{Thierry Donneau-Golencer}
\affil{Faros AI}

\renewcommand{\headeright}{}
\renewcommand{\undertitle}{}
\renewcommand{\shorttitle}{Repository-Grounded Rubrics}

\hypersetup{
pdftitle={Repository-Grounded Rubrics for Execution-Free Evaluation of Coding Agents},
pdfsubject={cs.SE, cs.LG},
pdfauthor={Matej Gerek, Danylo Zahorulko, Samuel Budai, Chris Rupley, Thierry Donneau-Golencer},
pdfkeywords={coding agents, software engineering evaluation, agentic rubrics, LLM-as-a-judge, execution-free evaluation, SWE-bench},
}

\begin{document}
\maketitle

\begin{abstract}
Coding agents are usually evaluated by running task-specific tests, but tests may be missing or incomplete, and running them requires a working environment for each repository.
We adapt agentic rubrics, introduced for selecting among candidate patches, into an execution-free measure of how far a patch satisfies a task's requirements.
An agent explores the repository before the change and writes weighted binary criteria, and a judge without tools scores a diff by the weighted share of criteria it satisfies.
We validate the ordering this score produces against test outcomes on SWE-bench Verified, using 7{,}291 labeled patches from fifteen systems. %
Within the 400 tasks that have both resolved and unresolved patches, the mean per-task AUROC is 0.826 (0.838 on tasks not used to tune the rubric-generation prompt), and ranking systems by their failing patches closely agrees with ranking them by resolve rate (Spearman 0.968). %
Case analyses trace disagreements to tests that do not match the task, execution-blocking defects that the textual assessment misses, and errors in criteria or verdicts.
The validation does not establish absolute score magnitudes or requirement satisfaction beyond agreement with tests, but rubric scores could complement tests where tests or execution environments cannot be assumed.
\end{abstract}


\section{Introduction}
\label{sec:introduction}

Coding-agent benchmarks such as SWE-bench evaluate patches by running task-specific tests.
In SWE-bench, a patch resolves a task when all designated tests pass \citep{jimenez2024swebench}.
This paper addresses settings where we cannot assume that tests are available, cover the task, and can be run.
The aim is to evaluate coding agents on the repositories they will actually work in, using tasks drawn from resolved issues and the code changes that addressed them.
Each task pairs an issue's problem statement with the repository state immediately before the corresponding change.
Each repository brings its own language, framework, and tooling, and with them its own requirements for an environment in which the code can be built and run.
Whether a task has tests, and whether those tests cover the change, is unknown in advance and varies from task to task.
Building and maintaining such an environment has a cost that is also unknown in advance and is paid for each repository.
The repository's history supplies one solution per task, but a task may admit many, so an evaluator cannot judge a patch solely by its resemblance to the recorded change.
We also need the outcome to say why a patch fell short, not only whether it did, because the consumer of an evaluation acts on the why.
Two patches can both fail to resolve a task while satisfying different parts of its requirements, so the evaluation should also measure partial satisfaction.
Taken together, these are the requirements: an evaluator that scores a patch without tests, without an execution environment, and without comparing it to a reference solution, that costs little per candidate patch, and that returns a graded, decomposable judgment.

Execution-based evaluation, reference-patch similarity, and direct language-model judging make different tradeoffs against these requirements.
Execution-based evaluation needs tests and an environment to run them in.
Benchmark tests can encode details of the reference solution and reject alternative implementations that satisfy the stated task \citep{openai2026verifiedretirement,openai2026proaudit}.
Although individual test failures provide diagnostic information, the aggregate pass/fail label does not describe which task requirements a patch satisfies.
Scoring a patch by its similarity to the recorded change needs no environment.
It measures resemblance to one solution, though, and fails every valid solution that looks different.
Direct language-model judging needs neither tests nor a reference, but a judge given only the issue text and diff cannot inspect repository code beyond the changed lines.

We use repository-grounded rubrics to evaluate patches under these constraints.
The framework has two stages: rubric generation and patch judging.
Before any patch is judged, an agent explores the repository at the commit preceding the change and writes a rubric for the task, a set of concrete, binary criteria with importance weights, each stating something a solution must do given the issue and the code as it stands.
A judge with no tools or repository access then marks each criterion for a candidate patch, and the score is the weighted fraction of criteria satisfied.
We selected the rubric-generation strategy prompt from a handwritten seed through automated prompt search using benchmark pass/fail labels.
The complete prompts and selection procedure are reported in \cref{app:optimization}.
We repurpose the agentic-rubric mechanism of \citet{raghavendra2026agenticrubrics}, introduced for selecting among candidate patches, as a measurement instrument, and evaluate its reliability and its ability to order patches within tasks on SWE-bench Verified.

We make three contributions.\footnote{Code and data: \url{https://github.com/faros-ai/repository-grounded-rubrics}.} %
First, we adapt agentic rubrics from selecting among candidate patches to measuring them.
We instruct the rubric generator to write criteria that any correct implementation can satisfy, define the score as an estimate of the weighted share of a task's requirements that a patch satisfies, and distinguish errors in the rubric's choice of criteria from errors in the judge's verdicts (\cref{sec:framework}).
Second, we present a protocol for validating a graded evaluator against binary ground truth and apply it across several repositories (\cref{sec:validation}).
The protocol measures repeatability across rubric generation and patch judging, scores on known-correct patches, how well the score separates resolved from unresolved patches within a task, and how closely a ranking of systems by the scores of their failing patches agrees with their ranking by resolve rate.
On SWE-bench Verified, the two system rankings closely agree (\cref{sec:capability}), and within tasks the score mostly ranks resolved patches above unresolved ones (\cref{sec:disc}).
The ranking result bears most directly on the main use we envisage, comparing models and agents on a team's own repositories (\cref{sec:applications}).
Third, we analyze how partial satisfaction, test defects and evaluator errors produce disagreement between rubric scores and test outcomes (\cref{sec:disagreement}).
\Cref{tab:scope} summarizes which claims the validation supports and which it does not establish. %

\section{Related work}
\label{sec:related-work}

We validate the rubric score on SWE-bench Verified \citep{openai2024swebenchverified}, a human-screened subset of SWE-bench \citep{jimenez2024swebench}.
Public submissions provide multiple labeled attempts at the same tasks, allowing us to assess whether the score ranks test-passing patches above failing ones within each task.
An OpenAI audit of a selected subset of difficult Verified tasks found tests that reject valid solutions by enforcing implementation details or checking functionality absent from the problem statement \citep{openai2026verifiedretirement}.
The same report documented evidence of training contamination and explained OpenAI's decision to stop reporting Verified scores for frontier models.
We discuss the risk of contamination in our generator and judge in \cref{sec:limitations}.
We examine how mismatches between tests and task requirements contribute to disagreement with rubric scores in \cref{sec:labelnoise}.

\citet{raghavendra2026agenticrubrics} introduced agentic rubrics for code.
An agent explores the repository and writes a checklist of weighted criteria for the task, after which a judge scores candidate patches against the checklist without running them.
They use the rubric as a verifier that selects the best of $K$ candidate patches, and evaluate it by the selected patch's resolution rate on SWE-bench Verified.
We adopt their agent-based generation, weighted binary criteria on four axes, and judging without repository access.
We use the resulting score as a measurement instrument.
We define what it estimates (\cref{sec:grs}), validate its ordering across the range of patch quality (\cref{sec:validation}), and retain the per-criterion verdicts for diagnostics (\cref{sec:applications}).
\citet{shi2025humanintheloop} also generate a per-bug rubric and use a language-model judge to measure patch validity against human consensus labels.
In contrast to our generator, theirs receives the ground-truth fix, and human developers revise the rubric before both human raters and the model judge apply it to patches from an internal repair system \citep{shi2025humanintheloop}.

Language-model judges are commonly assessed by their agreement with human preferences, and this work has also documented position and verbosity biases \citep{zheng2023judging}.
On coding tasks, execution-free pairwise judges can change their decision when candidate order is reversed, while independently scoring each candidate often yields tied scores \citep{jiang2026codejudgebench}.
On standalone functions, even the strongest model in one study frequently disagreed with test outcomes and changed correct verdicts after semantics-preserving rewrites \citep{crupi2025llmjudge}.
These studies evaluate the judge itself by comparing its verdicts with external labels.
We instead use the judge as one stage of a scoring instrument and validate the score's within-task ordering, repeatability, and behavior on reference patches (\cref{sec:validation}).

\section{Evaluation framework and score interpretation}
\label{sec:framework}

A rubric is a checklist of natural-language criteria that a correct patch is expected to satisfy, each with a weight reflecting its importance.
The framework consists of rubric generation followed by patch judging.
To generate the rubric, an agent explores the repository as it stood before the task was implemented and writes the criteria.
A judge then assesses each candidate patch against the criteria, marking each one as satisfied or not without using a test harness or repository-specific execution environment.
The mechanism follows the agentic rubrics of \citet{raghavendra2026agenticrubrics}, where it served as a verifier that selects the best of $K = 16$ candidate patches per task across the 500 tasks of SWE-bench Verified \citep{openai2024swebenchverified}.
We do not use it to select a patch.
We use the score as a measurement, so it has to mean something on its own, across tasks and across models, rather than only order candidates for the same problem.
This section specifies the mechanism (\cref{sec:instrument}), defines what the score measures (\cref{sec:grs}), and compares it to execution-based and other execution-free alternatives (\cref{sec:vs-tests}).

\subsection{Rubric generation and scoring}
\label{sec:instrument}

Throughout, a task $T$ is a problem statement together with the repository as it stood before the task was implemented, and $p$ is a candidate patch for it.

\paragraph{Rubrics.}
A rubric for $T$ is a finite set of criteria $R = \{(t_i, w_i, a_i)\}_{i=1}^{N}$.
Each criterion has a natural-language text $t_i$ stating one property a correct patch should have, a weight $w_i \in \{1,2,3\}$ marking it as nice-to-have, important, or must-have, and an axis $a_i$ selected from the four adopted from \citet{raghavendra2026agenticrubrics}.
File change concerns which files and symbols are edited, and how localized the edit is.
Spec alignment assesses whether the change does what the problem statement asks.
Integrity covers collateral damage such as test weakening, unrelated churn, or dependency changes.
Runtime concerns the behavior implied by the change.
Each criterion is atomic because it states one property, and static because that property can be assessed from the problem statement and the diff (\cref{sec:grs}).
The criteria are grounded in the repository and name the functions, types, and contracts the agent found there.

\paragraph{Rubric generation.}
The first stage deploys an agent into the repository at a commit preceding the task's implementation.
The agent works in a fresh copy of the repository at that commit, created without its version-control history, so later commits, including the reference solution, cannot be reached through git.
It therefore writes the rubric from the problem statement and the code as it stood before the task.

The agent receives the problem statement, the repository at that commit, a strategy prompt, and a size budget.
The size budget is the approximate number of criteria $N$ the rubric should contain.
It is derived from the reference patch, of which the agent learns only the line count.
We set it as
\begin{equation}
	\label{eq:size}
	N = \min\!\left(60,\ \operatorname{round}\!\left(8 + \frac{L}{12}\right)\right),
\end{equation}
where $L$ is the number of changed lines in the reference patch, added and removed lines counted alike.
The changed-line count serves as a rough proxy for task scope, allowing tasks with larger reference changes to receive more criteria.
This is the only use of the reference patch in rubric generation.
A setting without a reference patch would require a solution-independent sizing rule (\cref{sec:applications}).
The agent treats the budget as a soft target and distributes criteria across the four axes according to what the task requires, rather than padding or trimming the rubric to reach that number.
We set the constants after an exploratory comparison of three sizing rules on a development subset comprising all 44 \texttt{sphinx-doc/sphinx} tasks in SWE-bench Verified (\cref{app:sizing}). %
The rules produced rubrics of different lengths but similar mean per-task discrimination. %
We kept the rule whose count tracked the changed-line count most closely while producing the fewest criteria.

Given these inputs, the agent explores the repository and writes the rubric $R$, which the second stage uses to score candidate patches.
The harness around the agent (tools, output schema, axes, weight scale) is fixed; the strategy prompt was the only component varied during optimization.
Its main departure from \citet{raghavendra2026agenticrubrics} is that criteria must be satisfiable by any correct implementation, so they avoid fixing a location or an implementation path unless the problem statement requires it, and ancillary artifacts such as tests or changelog entries are capped at the lowest weight.

The generator uses fixed prompt templates adapted from \citet{raghavendra2026agenticrubrics} together with the separately supplied strategy prompt.
Starting from a handwritten strategy, we used GEPA \citep{agrawal2025gepa} to propose revisions using benchmark test verdicts as proxy feedback, rather than selecting every revision through manual iteration.
GEPA is a prompt optimizer that runs the system being optimized, reflects in natural language on the resulting trajectories and feedback to propose and test prompt revisions, and combines revisions from the candidates that perform best on at least one training example.
The selected strategy was then fixed for all rubric generation in the validation study.
The complete templates, deployed strategy, optimization objective, and selection procedure are reported in \cref{app:optimization}.

\paragraph{Judging.}
The second stage is a single model call with no tools and no access to the repository, with a fixed prompt, taken unchanged from \citet{raghavendra2026agenticrubrics}.
The judge receives the problem statement, the patch $p$ as a diff, and the criterion texts $t_i$; the weights and axes are withheld so that they cannot bias the verdicts.
For each criterion it returns a verdict $s_i(p) \in \{0,1\}$, and the rubric score is the weighted fraction of criteria satisfied,
\begin{equation}
	\label{eq:score}
	S_R(p) = \frac{\sum_{i=1}^{N} w_i\, s_i(p)}{\sum_{i=1}^{N} w_i} \in [0,1].
\end{equation}

\subsection{What the score measures}
\label{sec:grs}

\paragraph{Measurand.}
Let $Q(T)$ be the \emph{requirement population} of a task.
It contains every property of a patch that $T$ entails and that can be decided from the problem statement and the diff, with each property assigned an importance weight on the same three-point scale.
$Q(T)$ is never enumerated.
Its requirements and weights are those that a maintainer familiar with the repository would set for the task.
In this study no person sets them.
The generation agent estimates them, and the difference between its estimate and $Q(T)$ is the item-sampling error.
The quantity we want is
\begin{equation}
	\label{eq:grs}
	S^*(p) = \text{the weighted fraction of } Q(T) \text{ that } p \text{ satisfies},
\end{equation}
which we call \emph{grounded requirement satisfaction} because it measures satisfaction of requirements grounded in this repository and task.

The requirements in $Q(T)$ come from the task, independently of any particular solution.
Patches with no lines in common can therefore satisfy the same requirements and receive the same value of $S^*$.
A criterion that rejects a correct implementation because it takes a different approach does not belong in $Q(T)$.
The rubric-generation instructions described in \cref{sec:instrument} aim to exclude such criteria by avoiding implementation choices that the task does not require.
The generated rubrics do not fully achieve this ideal.
Reference-like resolved patches receive higher scores, and the difference is widest on the file-change axis (\cref{app:similarity}).

The scope of $Q(T)$ is limited to properties assessable from the problem statement and the diff without executing the code.
Failures that require additional runtime information to detect fall outside this scope, so a high score provides no assurance against them.
An unmet requirement evident from the supplied text falls within the scope of the assessment.
For example, a diff may change a function's signature while retaining calls that use the old signature.
If the score overlooks that incompatibility, the error lies in the rubric or the judge's assessment.
\Cref{sec:disagreement} examines both the limits of evaluation without execution and errors in the criteria and their assessment.

\paragraph{Instrument and estimator.}
The rubric $R$ is a finite \emph{sample} from $Q(T)$, drawn by the generation agent.
Because the agent reads the repository, its criteria are intended to reflect this task and codebase rather than general expectations about patches.
A criterion represents a requirement that a reviewer familiar with the repository would check.
The judge acts as a \emph{rater} whose verdict $s_i(p)$ estimates whether $p$ satisfies criterion $i$.
The reported score $S_R(p)$ is therefore an estimator of $S^*(p)$ with two error sources.
\emph{Item-sampling error} is the discrepancy between $R$ and $Q(T)$.
It includes requirements the agent omits and criteria outside $Q(T)$, such as ones that only a particular implementation can satisfy.
\emph{Rating error} is the gap between the judge's verdicts and the truth of the criteria.
The validation protocol measures variability across repeated judgments and repeated rubric generation.
Scores on reference patches and within-task discrimination assess their combined effects together with systematic rubric and rating errors (\cref{sec:validation}).

A score of $0.8$ means that the patch is estimated to satisfy 80\% of the importance-weighted requirements sampled for this task.
It is not a probability of passing tests, and it is not a percentage of the task completed.

A patch may pass the designated tests without satisfying every task requirement, and a high rubric score does not guarantee that it passes the tests.
The two measures should nevertheless agree in direction when the tests and rubric cover the same requirements.
We therefore validate the score by whether it orders resolved patches above unresolved patches within a task (\cref{sec:disc}).

Raw scores are comparable only among patches scored against the same rubric, because independently generated rubrics may sample different requirements.
Across rubric generations, we therefore compare patch rankings rather than score levels.

\subsection{Comparison with alternative evaluation methods}
\label{sec:vs-tests}

We compare the rubric framework with three alternatives that make different tradeoffs against the evaluation requirements defined in \cref{sec:introduction}.
These are execution-based evaluation, reference-patch similarity, and direct language-model judging.
The comparison assumes a fixed deterministic reference-similarity measure but does not prescribe how similarity is represented.
For the contamination check in \cref{app:similarity}, we use normalized line recall and token-trigram recall as two simple measures of whether a patch was written like the reference.
The direct judge receives the problem statement and diff without a rubric.

\begin{table}[t]
	\centering
	\footnotesize
	\renewcommand{\arraystretch}{1.15}
	\caption{Comparison of the rubric framework with three alternative approaches to scoring a patch.
		\emph{Tests} run the task's test suite, as in SWE-bench \citep{jimenez2024swebench}.
		\emph{Reference similarity} applies a fixed deterministic measure to the candidate and reference patches.
		\emph{Judge} is a model rating the patch from the problem statement and the diff alone, without a reference solution.
		CodeJudge \citep{tong2024codejudge} evaluates standalone functions this way, and the \emph{Judge} column assumes the same approach applied to repository patches.
		\emph{Rubric} is the instrument of \cref{sec:instrument}.
		Dollar costs are computed from token counts at the providers' prices when the runs took place, in August and September 2026 (\cref{app:cost}).}
	\label{tab:comparison}
	\newcommand{\rowrule}{\addlinespace[1.5pt]\cmidrule[0.03em]{1-5}\addlinespace[1.5pt]}
	\begin{tabular}{@{}>{\raggedright\arraybackslash}p{3.0cm}>{\raggedright\arraybackslash}p{2.7cm}>{\raggedright\arraybackslash}p{2.6cm}>{\raggedright\arraybackslash}p{2.8cm}>{\raggedright\arraybackslash}p{3.6cm}@{}}
		\toprule
		 & Tests & Reference similarity & Judge & Rubric \\
		\midrule
		Repository context & test suite and runtime & reference patch & problem statement and candidate diff & repository exploration during rubric generation \\
		\rowrule
		Measures & behavior on designated inputs & resemblance to reference patch & holistic model assessment & sampled task requirements \\
		\rowrule
		Depends on & test suite, environment & reference patch and fixed measure & judge model & generator, strategy, and judge model \\
		\rowrule
		Needs an execution environment & yes & no & no & no \\
		\rowrule
		Needs existing tests & yes & no & no & no \\
		\rowrule
		Needs a reference patch & no & yes & no & its line count only \\
		\rowrule
		Setup cost per task & a suitable execution environment & none & none & one agent run, \$0.25 at the median \\ %
		\rowrule
		Cost per candidate & a test run & one similarity computation & one model call & one model call, half a cent at the median \\ %
		\rowrule
		Source of run-to-run variation & tests and environment & none & judge model & generator and judge models \\
		\rowrule
		Verdict detail & per-test outcomes & measure-specific matches & not requested in this comparison & per-criterion outcomes with text and weight \\
		\rowrule
		Graded output & fraction of tests passed & similarity score & model rating & weighted fraction of criteria satisfied \\
		\bottomrule
	\end{tabular}
\end{table}

Tests provide direct evidence about behavior on the inputs they exercise.
Their use requires suitable tests that cover the requested change and a repository-specific environment in which to run them.
The environment may be shared across tasks or configured for a particular task and repository state.
Some benchmark implementations incur per-instance setup, such as sandbox initialization \citep{raghavendra2026agenticrubrics}.
Our setting does not assume that tests and an environment are available.
Even when suitable tests and an environment are available, tests may encode choices made by one solution, such as a name, a signature, or the exact text of an error message.
A correct patch that takes a different route can fail such a test, and a measure of resemblance to one solution inherits that solution's choices.
\Cref{sec:disagreement} illustrates this source of disagreement between the score and the tests.
The rubric generator does not see the content of the reference patch or any candidate patch, and its strategy instructs it to avoid implementation choices absent from the task.
Agent-generated tests, which \citet{raghavendra2026agenticrubrics} also evaluate, are an alternative when a suitable execution environment is available.
Both generated tests and rubric criteria can encode assumptions absent from the task, but generated tests must additionally be executed before they yield a verdict.

Test-based and rubric-based evaluation measure different properties.
In benchmarks such as SWE-bench, the task-level test verdict is binary because a patch resolves a task only when all designated tests pass \citep{jimenez2024swebench}.
A fraction of tests passed could provide a graded output, but its granularity depends on the suite, and the SWE-bench task-level verdict does not weight tests by the importance of the requirements they check.
The rubric score is graded by construction from weighted criteria sampled from the task's requirements, and it can be computed without execution.
Disagreement can arise from differences in coverage as well as errors in the tests, rubric, or judge (\cref{sec:disagreement}).
The rubric complements rather than replaces tests.
It assesses requirements visible in the problem statement and diff, including scope and collateral changes, and reports which sampled requirements the patch satisfies.
It can be applied without a repository-specific execution environment.
Unlike an executed test assertion, a rubric verdict is a model judgment and therefore introduces rating variability, which we measure in \cref{sec:repeatability}.

\section{Validation protocol and results}
\label{sec:validation}

The weighted rubric score estimates grounded requirement satisfaction and is subject to item-sampling and rating error (\cref{sec:grs}).
We assess repeatability across rubric generation and patch judging, scores on reference patches, and within-task discrimination between resolved and unresolved patches.

\subsection{Validation protocol}
\label{sec:protocol}

We validate against SWE-bench Verified \citep{openai2024swebenchverified}.
\citet{raghavendra2026agenticrubrics} also used this benchmark to evaluate agentic rubrics for candidate selection.
The benchmark's public leaderboard submissions provide multiple labeled patches per task across several repositories, which a within-task ranking metric requires.
A patch receives a resolved label if it passes every designated test.
The audit cited in \cref{sec:related-work} found tests that reject valid solutions, so some of these labels may be wrong.
\Cref{sec:labelnoise} treats such cases as one source of disagreement between the score and the tests.
A boolean label can tell us whether a higher score means a more likely resolved patch.
It cannot tell us whether a score of 0.7 is closer to the truth than 0.6, because it has no notion of closeness.
Validation therefore concerns direction at the pass/fail boundary, and magnitude is left unvalidated (\cref{sec:grs}).

We assess repeatability at both stages of the instrument (\cref{sec:repeatability}).
Repeated judgments of the same patches against fixed rubrics isolate variability in the judge's verdicts.
Independent rubric generations expose changes in the sampled criteria together with residual judgment variability.
Agreement in scores and within-task orderings describes the effect of each source, but does not establish whether the criteria or verdicts are correct.

Scores on reference patches help assess possible over-specification.
We report their distributions overall and by axis (\cref{sec:gold}).
A failed criterion may reflect an unnecessary requirement, a shortcoming of the reference patch, or an incorrect judgment.
The distributions cannot distinguish these causes, and this check does not measure requirements omitted from the rubric.

Per-task AUROC, the area under the ROC curve computed within each task, measures discrimination between resolved and unresolved patches.
For a task with both a resolved and an unresolved attempt, it is the probability that the rubric scores a uniformly drawn resolved attempt above a uniformly drawn unresolved one, ties counted as half, which is the Mann--Whitney $U$ statistic normalized by the number of resolved--unresolved pairs.
We average it over such tasks with equal weight, report the across-task distribution in \cref{sec:disc}, and give a 95\% percentile bootstrap interval over those tasks (2,000 resamples), which covers the sampling of tasks and not the sampling of attempts within a task.\footnote{A bootstrap interval is obtained by redrawing the set of tasks at random with replacement many times, recomputing the mean AUROC on each redraw, and taking the central 95\% of the resulting values. It describes how much the mean would move under a different sample of tasks from the same benchmark.} %
Tasks whose attempts all resolve or all fail contain no such pair and are excluded.
The statistic therefore applies only to tasks on which the evaluated system pool produced both outcomes, and we report the number of these contested tasks alongside each AUROC.
We also plot the score distributions of the resolved and unresolved classes.

\subsection{Validation population and instrument configuration}
\label{sec:populations}

The validation population is the whole of SWE-bench Verified, 500 tasks from twelve Python repositories, each with its reference (gold) patch and its tests. %
The patches we score are the public submissions of fifteen systems on the SWE-bench leaderboard, listed in \cref{tab:systems} and described in \cref{tab:system-configurations}, together with the harness verdict each system published for them. %
Of the 7{,}500 cells in the grid of tasks by systems, 201 are empty because the system produced no patch for the task or published no evaluation log, which leaves 7{,}299 labeled patches, 3{,}688 of them resolving their task. %
Eight submitted patches exceed the judge's context window, six because they include generated build artifacts and two because they bundle a large installation script. %
They remain in the population but have no rubric score.
Within-task statistics require a task with at least one resolved and one unresolved patch, a condition that 400 of the 500 tasks meet. %
The remaining 35 tasks were resolved by every system and 65 by none, and they contribute to the reference-patch and class distributions but not to the ordering statistics. %
\Cref{tab:populations} gives the counts per repository.

Each task has a single rubric, generated once by the rubric-generation agent described in \cref{sec:instrument} with claude-sonnet-4-6 under the deployed strategy of \cref{app:optimization} and sized by \cref{eq:size}. %
The reference patches have a median of 7 changed lines (p90 33, maximum 232) in a median of one file (p90 2, maximum 21). %
The 500 rubrics have a median of 10 criteria, ranging from 8 to 24, and their length follows the changed lines of the reference patch with a Pearson $r$ of 0.90 (\cref{app:population}).\footnote{Pearson's $r$ measures linear association between two numerical variables, ranging from $-1$ for a perfect negative linear relationship to 1 for a perfect positive one, with 0 indicating no linear relationship.} %
Each patch was judged once by gpt-5.4 using the provider's default settings with no reasoning effort. %
We treat this configuration as part of the instrument because \cref{sec:repeatability} shows that changing the reasoning effort changes the scores. %
All numbers in \cref{sec:gold,sec:disc} come from this one configuration.
The repeatability study repeats both stages on a 100-task subset of this population.

Three scored patches have no published verdict and are excluded from every statistic that needs a label. %
Of the 500 tasks, 85 were instances in the optimization of the strategy prompt (\cref{app:optimization}), 66 in its training split and 19 in its validation split, so the discrimination results are reported on all 500 tasks and, beside them, on the 415 tasks the optimizer never saw. %

\begin{table}[t]
	\centering
	\footnotesize
	\renewcommand{\arraystretch}{1.1}
	\caption{The validation population comprises SWE-bench Verified submissions from fifteen leaderboard systems. \emph{Contested} tasks have at least one resolved and one unresolved patch and are the tasks the ordering statistics of \cref{sec:disc} rest on.}
	\label{tab:populations}
	\begin{tabular}{@{}lrrrr@{}}
		\toprule
		Repository & Tasks & Contested & Patches & Resolved \\
		\midrule
		django & 231 & 188 & 3,370 & 1,781 \\
		sympy & 75 & 59 & 1,109 & 524 \\
		sphinx & 44 & 35 & 638 & 228 \\
		matplotlib & 34 & 27 & 461 & 218 \\
		scikit-learn & 32 & 27 & 477 & 345 \\
		astropy & 22 & 15 & 327 & 133 \\
		xarray & 22 & 17 & 328 & 176 \\
		pytest & 19 & 17 & 281 & 154 \\
		pylint & 10 & 5 & 148 & 41 \\
		requests & 8 & 8 & 116 & 73 \\
		seaborn & 2 & 1 & 30 & 2 \\
		flask & 1 & 1 & 14 & 13 \\
		\midrule
		all & 500 & 400 & 7,299 & 3,688 \\
		\bottomrule
	\end{tabular}
\end{table}

\subsection{Repeatability}
\label{sec:repeatability}

The framework has two stochastic stages, rubric generation and patch judging.
We measure their repeatability separately because a fixed-rubric study does not include variation in the criteria selected for a task.

To isolate variability in patch judging, we scored each patch in the study five times against a fixed rubric under the production configuration. %
The study uses a manifest of 100 SWE-bench Verified tasks, fixed before any judge call. %
It contains the 44 \texttt{sphinx-doc/sphinx} tasks of the sizing comparison (\cref{app:sizing}) and 56 tasks drawn at random from the other repositories in proportion to their number of tasks with at least three labeled submissions, excluding tasks used in strategy optimization. %
For each task, we selected the lowest-, median- and highest-scoring non-reference submissions using their original scores, together with the reference patch, giving 400 patches and 2{,}000 judge calls. %

The mean Pearson correlation between runs is 0.985, and the mean within-patch standard deviation is 0.020. %
Among non-reference patches, the mean standard deviation is 0.012 for original scores in $[0.8,1.0]$ and 0.034 for scores in $[0.4,0.6)$. %
Criterion verdicts differ across the five runs for 4.8\% of patch--criterion pairs, while 275 of the 400 patches receive identical aggregate scores on every run. %
The mean within-task Kendall $\tau_b$ across run pairs is 0.960, with the ordering of all four patches reproduced exactly across every run pair for 63 of the 100 tasks.\footnote{Kendall's $\tau_b$ measures agreement between rankings while accounting for ties, ranging from $-1$ for reverse orderings to 1 for identical orderings, with 0 indicating no ordinal association.} %
This variability warrants caution when interpreting small differences between single-run scores, but does not establish a universal threshold for distinguishing patches.

With reasoning effort set to \emph{medium} on the same manifest, pooled repeatability decreases.
The mean run-pair correlation falls to 0.964, the mean within-patch standard deviation rises to 0.030, and criterion verdicts differ across runs for 6.9\% of patch--criterion pairs. %
The mean within-task Kendall $\tau_b$ falls to 0.923, while completion-token use increases approximately tenfold. %
Changing the effort setting also changes the scores, with five-run mean scores differing by more than 0.10 for 107 of the 400 patches. %
These results support retaining no reasoning effort for repeatability in this configuration, without establishing which setting is more accurate.

To measure repeatability across generated rubrics, we independently generated three rubrics for each task in the same 100-task manifest under the production configuration (\cref{app:repeatability}).
For the 1{,}451 labeled leaderboard submissions to these 100 tasks, scored under every generation, scores assigned to the same patch differ by about 0.11 on average between generations. %
On the four-patch sample, pairwise within-task Kendall $\tau_b$ ranges from 0.732 to 0.750 across generations, compared with 0.960 across repeated judgments against fixed rubrics. %
The mean task-level standard deviation is 0.0814 between rubric generations and 0.0200 within a fixed rubric, an observed ratio of 4.06. %
Aggregate discrimination varies less, with mean per-task AUROC from 0.818 to 0.842 across generations on the 77 contested tasks in this subset. %
Judge variability measured with a fixed rubric therefore applies only to patches scored against the same rubric.
Regenerating the rubric adds a larger source of variation in scores and task-level orderings.
Comparisons between patches scored against different rubrics, and conclusions resting on a single rubric, inherit this variation.

\subsection{Scores on reference patches}
\label{sec:gold}

\begin{figure}[t]
	\centering
	\includegraphics{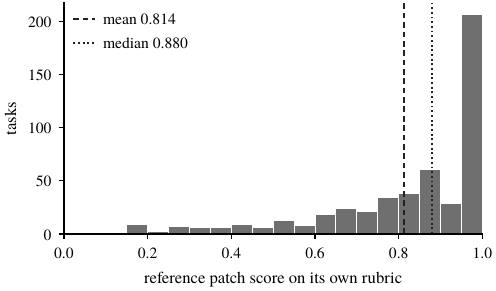} %
	\caption{Score of each reference patch on its own rubric, 500 SWE-bench Verified tasks. The dashed line marks the mean (0.814) and the dotted line the median (0.880). %
	}
	\label{fig:gold}
\end{figure}

Each of the 500 reference patches was scored against the rubric written for its task, and \cref{fig:gold} shows the resulting distribution. %
The mean score is 0.814 and the median 0.880, with a quarter of the reference patches below 0.739 and a quarter above 0.957. %
Fifty-nine reference patches (11.8\%) satisfy every criterion, while 47 (9.4\%) score below 0.5. %
Reference patches on the 415 tasks outside the optimization set average 0.821, and those on the 85 inside it average 0.775. %
We report the distribution rather than a pass count, because the instrument defines no pass threshold.

Broken down by axis, the reference patches lose most on file-change (mean 0.750) and least on integrity (0.883), with spec alignment at 0.819 and runtime at 0.834, and file-change is the lowest of the four axes for 265 of the 500 reference patches. %
A criterion that the reference patch fails may be a requirement the task does not impose, which is the over-specification of \cref{sec:grs}, a shortfall of the reference patch itself, since the merged patch is one correct solution rather than a perfect one, or an error of the judge.
The overall distribution and differences among axes do not separate these three causes, so they cannot estimate the prevalence of over-specification or explain why scores differ by axis.

\subsection{Within-task discrimination}
\label{sec:disc}

Per-task AUROC is defined only for tasks with patches of both classes, which excludes the 35 tasks resolved by every system that attempted them and the 65 resolved by none. %
Over the remaining 400 contested tasks and their 5,834 patches, the mean per-task AUROC is 0.826 with a 95\% interval of [0.805, 0.847]. %
The interval reflects the sampling of tasks only and does not include variation from repeated judgments or regenerated rubrics (\cref{sec:repeatability}).
Weighting each task by its number of resolved--unresolved pairs, which ranges from 12 to 56 with a median of 36, gives 0.836 [0.818, 0.854], so tasks with few pairs do not inflate the unweighted mean. %
The distribution behind the mean is skewed, with 109 of the 400 tasks (27\%) ordered perfectly and 35 (9\%) at or below chance, which places the median at 0.901. %
The statistic therefore applies only to tasks on which the evaluated system pool produced at least one patch of each class.
It says nothing about the tasks that every system solves or none does.

\begin{figure}[t]
	\centering
	\includegraphics{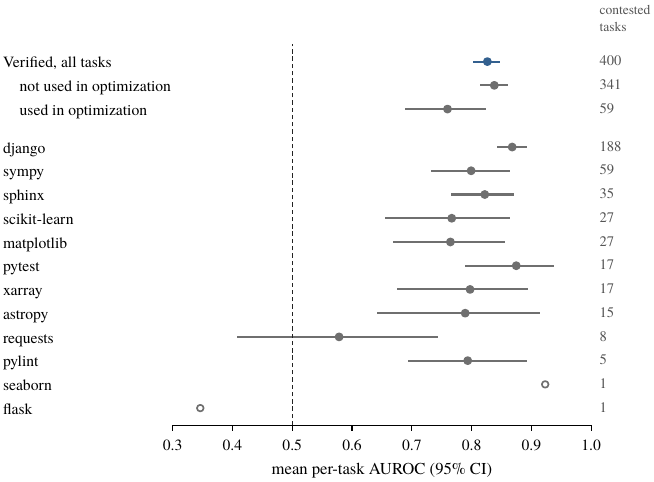} %
	\caption{Mean per-task AUROC on SWE-bench Verified, resolved above unresolved within a task, with 95\% bootstrap intervals over tasks. %
	Numbers at the right count contested tasks, which have at least one patch of each class and are the only ones the statistic is defined on.
	Repositories are ordered by that count.
	Seaborn and flask have one contested task each and no interval (open circles).
	The dashed line marks chance.
	\Cref{tab:auc} lists the values.}
	\label{fig:auc}
\end{figure}

Because 85 of the tasks were training or validation instances of the strategy optimization in \cref{app:optimization}, \cref{fig:auc} also shows the 415 tasks the optimizer never saw, where the mean per-task AUROC is 0.838 [0.817, 0.859], against 0.760 [0.691, 0.822] on the 85. %
The repository-level estimates are descriptive, and some rest on few contested tasks.

On this population, changing the sizing rule produced little change in mean per-task AUROC.
On the 44 sphinx tasks, three rubric sets sized by different rules, with 12.3, 22.8 and 10.4 criteria per rubric on average, gave per-task AUROC of 0.754, 0.754 and 0.751 over the same 32 contested tasks and 545 patches (\cref{app:sizing}). %
Those rubric sets were generated with an earlier configuration than the one evaluated in this section, and the comparison covers one repository of small changes.

\begin{figure}[b]
	\centering
	\includegraphics{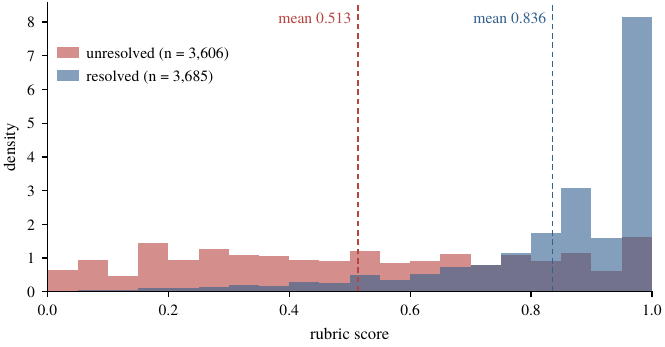} %
	\caption{Score distributions of resolved and unresolved patches, 7,291 labeled patches over 500 SWE-bench Verified tasks. Dashed lines mark the class means (0.836 and 0.513), which are the thresholds for leniency and strictness. %
	}
	\label{fig:classes}
\end{figure}

\Cref{fig:classes} complements the within-task AUROC with pooled score distributions for resolved and unresolved patches, including disagreements analyzed in \cref{sec:disagreement}.
Resolved patches average 0.836 (median 0.885, SD 0.182) and unresolved patches 0.513 (median 0.500, SD 0.286), a gap of 0.32. %
We call an unresolved patch scoring at or above the resolved-class mean a case of \emph{leniency}, and a resolved patch at or below the unresolved-class mean a case of \emph{strictness}.
The figure shows 645 leniency cases among the 3,606 unresolved patches (17.9\%) and 289 strictness cases among the 3,685 resolved patches (7.8\%). %

Scored one axis at a time, three of the four axes discriminate about equally while the fourth barely does. %
Per-task AUROC on the file-change axis alone is 0.773 [0.751, 0.793], on spec alignment 0.766 [0.744, 0.786], on runtime 0.760 [0.740, 0.778], and on integrity 0.583 [0.567, 0.598]. %
Integrity checks for test weakening, mass renames, unnecessary dependency changes, and changes to public interfaces that the task does not require.
A patch can avoid these problems without fixing the reported issue, so satisfying integrity criteria does not necessarily distinguish a successful fix from an unsuccessful one.

The generator model was trained after every benchmark task was published, so agreement with the tests could in principle depend on resemblance to a remembered reference patch.
\Cref{app:similarity} compares resolved patches written like the reference with those written differently.
The rubric separates differently written patches from failures at a per-task AUROC of 0.763 [0.733, 0.791], compared with 0.872 [0.849, 0.893] for reference-like patches. %
Reference-like resolved patches score higher, especially on the file-change axis, but discrimination persists among patches written differently from the reference.

\subsection{Ranking systems from failing patches}
\label{sec:capability}

\begin{figure}[t]
	\centering
	\includegraphics{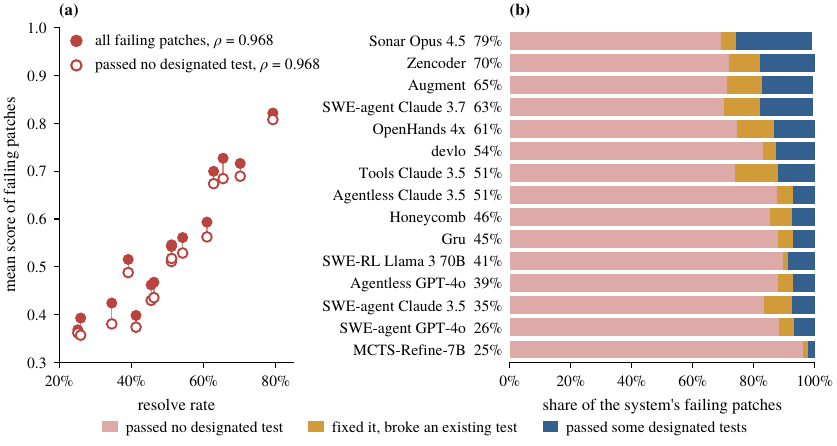} %
	\caption{Failing patches by system on SWE-bench Verified, 15 systems ordered by resolve rate. (a) Mean rubric score of a system's failing patches against its resolve rate, over all 3,606 failing patches (filled) and over the 2,999 that passed none of the tests they had to make pass (hollow). %
	(b) Each system's failing patches grouped by whether they passed none of the designated tests, passed all of them but broke a test that had passed before, or passed some. %
	}
	\label{fig:capability}
\end{figure}

A system's mean score over all its patches would largely restate its resolve rate, because resolved patches score higher than unresolved ones (\cref{sec:disc}) and score similarly across systems.
Tests give every failing patch the same label, so failing patches are where the score can carry information about a system that its resolve rate does not.
On SWE-bench Verified, mean rubric scores on failing patches produce a system ranking that closely agrees with the ranking by resolve rate on this pool.
\Cref{fig:capability}a plots each system's resolve rate against its mean score on those patches.
The Spearman rank correlation between the rankings is 0.968 [0.928, 0.971] over the fifteen systems and 3,606 failing patches.\footnote{Spearman rank correlation measures agreement between two rankings, ranging from $-1$ for reverse rankings to 1 for identical rankings, with 0 indicating no rank association.} %
The correlation remains when scores are compared only among failing patches on the same task (0.950) and when only the 42 tasks that every system failed are used (0.914). %
These controls address differences in which tasks each system fails.
The result establishes agreement between system rankings on this pool.
It does not establish the correct ordering of every pair of failing patches.

Resolved patches have mean scores between 0.766 and 0.884 across systems, while the mean scores of failing patches rise with resolve rate. %
Consequently, the gap between a system's passing and failing patches narrows from 0.480 for the weakest system to 0.062 for the strongest. %
This is the same association as the failure-based ranking, viewed through the separation between the classes.
A high score on a failing patch is therefore more common among the more successful systems, rather than a discrepancy distributed uniformly across them.

The per-test records provide a separate comparison with graded test outcomes among unresolved patches.
An unresolved patch can pass none of the tests it had to make pass, some of them, or all of them while breaking a test that had passed before.
\Cref{fig:capability}b shows the share of failures in the latter two groups rising from 3.8\% for the weakest system to 29.8\% for the strongest, with a Spearman correlation of 0.904 [0.686, 0.914] with resolve rate. %
Patches passing no target tests score 0.474 on average, those passing all target tests but introducing a regression score 0.684, and those passing some target tests score 0.719. %
The scores thus distinguish groups with different test records even though all receive an unresolved label.
The association also persists where none of the target tests pass, shown by the hollow points in \cref{fig:capability}a, with a Spearman correlation of 0.968 between mean score and resolve rate. %
Within that group, however, the test records do not distinguish the patches and therefore cannot check the score differences.

\section{Interpreting disagreement with tests}
\label{sec:disagreement}

The overlapping score distributions in \cref{sec:disc} include unresolved patches with high rubric scores and resolved patches with low ones.
Neither pattern alone establishes an error in the rubric score, which estimates requirement satisfaction rather than the probability of passing tests (\cref{sec:grs}).
A patch can satisfy many requirements while missing a condition necessary for the tests to pass, and a passing patch can omit requirements the tests do not check.
Disagreement can also arise from tests that misrepresent the task, defects missed by evaluation without execution, or errors in the criteria and their assessment.

\subsection{Mismatch between tests and task requirements}
\label{sec:labelnoise}

A test can reject a valid solution by enforcing an implementation choice or by checking functionality absent from the task description.
OpenAI distinguishes these as narrow and wide tests in its audit of SWE-bench Verified \citep{openai2026verifiedretirement}.
The audit examined 138 tasks that o3 did not solve consistently and found material issues in the tests or problem descriptions of 59.4\% of them. %
These tasks are a selected hard subset, so the rate cannot be applied to the benchmark as a whole or to our disagreement cases.

For narrow tests, OpenAI gives \texttt{pylint-4551}, whose issue asks for Python type hints to appear in UML diagrams.
The tests import a helper named \texttt{get\_annotation}, although the issue specifies no such function.
An implementation can provide the requested behavior without that helper and fail when the tests import it.
For wide tests, the audit gives \texttt{sympy-18199}, whose task description concerns a missing zero root in \texttt{nthroot\_mod}.
The associated reference patch and tests also address other issues that the task description does not include, so implementing the described fix alone need not pass the tests \citep{openai2026verifiedretirement}.

In \texttt{django-16667}, the test enforces an unstated return value.
The issue reports a crash on an out-of-range year.
The candidate catches the error in the existing handler, returning the fallback string already used for other malformed input.
The reference instead returns the literal \texttt{"0-0-0"}, and the tests assert that literal, which the issue never mentions.
Thirteen of the fifteen submitted patches add \texttt{OverflowError} to the existing handler, and the nine of them that modify only \texttt{widgets.py} score 1.0. %
None of the fifteen resolves the task. %
The rubric credits the alternative handling of the reported crash, while the test rejects its return value (\cref{app:cases}).

Tests can also accept a patch that violates a requirement they do not check.
In \texttt{django-14122}, the issue asks that \texttt{Meta.ordering} fields stop being added to \texttt{GROUP BY}.
The candidate removes the mechanism that adds order-by expressions to the grouping, so expressions introduced only by an explicit \texttt{order\_by()} are no longer added either.
The test designated to change from failing to passing uses only \texttt{Meta.ordering} and passes, as do the designated regression tests.
The rubric includes a criterion requiring explicit ordering to retain its contribution to grouping, and the judge marks it unmet.
The patch scores 0.083 against 0.958 for the reference. %
The explicit-ordering criterion identifies a regression that the designated tests miss.
Most of the gap to the reference comes from other criteria, some of which prescribe the reference's mechanism or mark behavior the candidate preserves as unmet (\cref{app:cases}).

\subsection{What evaluation without execution can miss}
\label{sec:execution-limit}

The framework avoids the per-repository build and execution setup described in \cref{sec:introduction}.
Its behavioral judgments therefore rest on reading the diff against the criteria, without an independent check that the patched code builds, imports or runs.
In \texttt{astropy-8707}, a candidate adds bytes handling to the requested methods but also inserts a malformed line into another method.
The patch nevertheless scores 0.943 despite the malformed hunk and resulting import-time harness error (\cref{app:case-astropy8707}). %

Some execution-blocking defects, including this indentation error, can be detected by static inspection or a syntax checker.
A language-specific syntax check can be reused across repositories without a complete execution environment.
Broader checks can require the repository's dependencies, build configuration or runtime environment, and some failures depend on inputs or conditions absent from the problem statement and diff.
Establishing behavior under those conditions requires evidence beyond the textual assessment, such as a build or an execution in the relevant environment.
The framework avoids requiring that setup for each repository, at the cost of providing no independent assurance of executability.

The Astropy example is a visible defect the evaluation missed, within the scope of \cref{sec:grs}; it does not show that the defect required execution to detect.
It illustrates how a high score can coexist with an execution-blocking error when evaluation relies on the sampled criteria and the judge's reading alone.

\subsection{Rubric and judge errors}
\label{sec:instrument-share}

The absence of execution is one way an assessment can miss a defect, but disagreement can also arise from an incorrect verdict on the code the judge sees.
In \texttt{django-10914}, the issue asks for a default of \texttt{0o644} for \texttt{FILE\_UPLOAD\_PERMISSIONS}.
Twelve of the fourteen systems that submitted a patch make the same effective settings change as the reference. %
The reference satisfies the criteria concerning the default permissions, preservation of an explicit override and the ability to opt out, all decided by that settings line.
The judge nevertheless returns different verdicts for patches with this same effective change.
Among the four patches with no other changes, three score 0.76 and one scores 0.24. %
A candidate that also changes documentation scores 0.143. %
\Cref{app:cases} shows the criterion verdicts and the changes accompanying each patch.
This is rating error in the terms of \cref{sec:grs}.
The repeated-judgment study does not determine which verdict is correct (\cref{sec:repeatability}).

Errors can also originate in the criteria rather than their assessment.
A criterion may require an implementation detail that the task leaves open, reducing the score of an otherwise valid alternative.
The reference-similarity analysis in \cref{app:similarity} finds that resolved patches resembling the reference receive higher scores than those written differently.
That association is consistent with overly specific criteria, but does not identify which criteria are invalid or establish how much of the score difference they cause.
Likewise, a criterion failed by the reference can reflect an unnecessary requirement, a shortcoming of the reference, or an incorrect judgment (\cref{sec:gold}).

A rubric can also omit a task-specific requirement, leaving a patch's violation of it unaccounted for even when the written criteria are assessed correctly.
The present measurements do not establish how often requirements are omitted or how often individual verdicts are wrong.

Taken together, the cases show why a disagreement must be interpreted against the task and the evidence supporting each verdict.
A failing test can expose a genuine defect or enforce something the issue did not require.
A low rubric score can identify an unchecked defect or reflect an error in the criteria or judgment.
The examples establish these mechanisms in particular cases, without attributing population shares to them.

\section{Applications}
\label{sec:applications}

Choosing a coding agent requires evaluating it on tasks representative of the repositories where it will be used.
Public benchmarks may not reflect those tasks, while building an evaluation around repository-specific tests requires suitable tests and an environment in which to run them.
Our results suggest that rubric scores could support these comparisons.
On SWE-bench Verified, system rankings based on scores of failing patches closely agree with rankings based on resolve rates (\cref{sec:capability}). %

A repository-specific benchmark could draw tasks from resolved issues and the code changes that addressed them.
A rubric would be generated for each task using the problem statement and the repository at the commit preceding the change.
The same rubric could then score patches from different models or harnesses without running tests or consulting the reference solution at scoring time.
The repository checkout, rubric generation, and scoring procedure could be reused without building a separate execution environment for each repository.
The framework could also support private benchmarks whose tasks and solutions were unavailable during model training, avoiding the corresponding exposure risk of public benchmarks.

Rubric generation is a multi-step agent run and costs about \$0.25 per task at the median with the production generator, claude-sonnet-4-6, and prompt caching enabled, at list prices of September 2026 (\cref{app:cost}). %
Judging is a single model call and costs about half a cent per candidate at the median (\cref{app:cost}). %
Once rubrics exist for a task set, scoring patches from a second harness, a new model version, or a best-of-$N$ sweep adds only the judging cost, a fraction of a cent per candidate at the median. %
Execution-based scoring instead requires per-repository setup and a test run for every candidate.
Judging cost follows the length of the patch more than the size of the rubric. %
Within each quartile of patch length, rubrics of twelve or more criteria cost between a tenth and a fifth of a cent more per judgment than rubrics of nine or fewer (\cref{app:cost}). %

A judged rubric is a set of natural-language propositions about the patch, each marked satisfied or not, so its output can be aggregated where a test result cannot.
Criteria marked unmet could be grouped by module, task type, or patch size to identify recurring patterns in the judge's assessments across models and over time.
Such analyses could suggest patterns of the form \emph{on bug fixes over 100 lines in the billing module, the agent solves the issue but tends not to preserve adjacent behavior}, or \emph{the latest model upgrade delivers the requested behavior more reliably but has become worse at updating call sites}.
Grouping criterion verdicts by axis could help identify whether a model repeatedly falls short on specification alignment, edit scope, integrity, or runtime behavior.

Comparisons across task categories could inform model selection.
For example, if a cheaper model receives similar rubric scores to a frontier model on small bug fixes in a particular module, it could be a candidate for handling those tasks at lower cost.

Nothing in rubric generation requires the task to have been implemented, so the same procedure could produce grounded acceptance criteria before any code is written, a definition of done at planning time.
The size budget in \cref{eq:size} would therefore need to be set by a solution-independent rule.
Failing criteria could also serve as feedback to the agent that produced the patch, and our preliminary trials of such a repair loop are encouraging, improving resolution on some repositories and leaving it unchanged on the others. Beyond those trials, neither use has been tested.

\section{Limitations and future work}
\label{sec:limitations}

\Cref{tab:scope} summarizes which claims the validation supports and where each is examined.
The rest of this section discusses the claims it does not establish.

\begin{table}[t]
	\centering
	\small
	\renewcommand{\arraystretch}{1.15}
	\caption{Scope of the validation. \emph{Supported} claims are validated against test outcomes on SWE-bench Verified, \emph{partly supported} claims rest on limited or indirect evidence, and the remaining claims were not established or not assessed.}
	\label{tab:scope}
	\begin{tabular}{@{}>{\raggedright\arraybackslash}p{6.4cm}>{\raggedright\arraybackslash}p{6.6cm}l@{}}
		\toprule
		Claim & Status & Where \\
		\midrule
		Scores repeat across judgments with the same rubric & Supported. They vary more across regenerated rubrics & \S\ref{sec:repeatability} \\
		Within a task, resolved patches score above unresolved ones & Supported, including on tasks excluded from strategy optimization & \S\ref{sec:disc} \\
		Ranking systems by their failing patches agrees with ranking by resolve rate & Supported on 15 systems of one benchmark & \S\ref{sec:capability} \\ %
		Discrimination does not depend on recall of the reference solution & Partly supported. It holds for patches written differently from the reference, but recall cannot be ruled out & App.~\ref{app:similarity} \\
		Criteria are legitimate and complete, and verdicts are correct & Partly supported, by case analyses only & \S\ref{sec:instrument-share} \\
		The score measures requirement satisfaction, including distances between scores, beyond its agreement with tests & Not established & \S\ref{sec:protocol} \\
		A patch builds and runs & Not assessed & \S\ref{sec:execution-limit} \\
		Results hold beyond Python, SWE-bench Verified and one generator and judge configuration & Not established & \S\ref{sec:populations} \\
		\bottomrule
	\end{tabular}
\end{table}

The evaluated tasks come from Python repositories, and their reference patches typically change few lines in one file. %
All validation rubrics were produced by one generator model and strategy, and every candidate was scored by one judge configuration (\cref{sec:populations}).
We did not systematically compare generator models, judge models, or rubric-generation harnesses.
The multilingual tasks used during strategy optimization provide development provenance but do not establish validity outside Python.

The assessment also has no independent evidence that a patch builds, runs, or behaves correctly under conditions that cannot be inferred from the problem statement and diff (\cref{sec:grs,sec:execution-limit}).
Some execution-blocking defects can be detected without running the code, but the framework does not supplement its task-specific rubrics with independent static checks.
Criterion quality and judgment accuracy remain only partly assessed.
The present evaluation does not measure how often a rubric omits a requirement or includes a criterion that a valid implementation need not satisfy.
Repeatability across judgments and rubric generations measures consistency, but not whether the criteria are legitimate and complete or whether their verdicts are correct (\cref{sec:repeatability,sec:instrument-share}).
\citet{shi2025humanintheloop} report lower agreement when their judge uses an unrevised generated rubric instead of the human-refined rubric used by their human raters.
Our validation does not test whether a comparable review step would improve our criteria.
Human review of generated criteria and individual verdicts could assess criterion legitimacy and rating accuracy.
Measuring omissions would additionally require experts to identify requirements independently of the generated rubric, since reviewing existing criteria cannot reveal what the rubric failed to include.

The strategy optimization and model training histories create two separate dependencies on public benchmark data.
The strategy was optimized against test labels under an objective that was not derived from the definition of grounded requirement satisfaction (\cref{eq:optimization-fitness,sec:grs}).
It rewards a test-passing patch for reaching $S_R(p)=1$, gives the same fitness to every failing patch below 1, and penalizes a failing patch at 1 even when the failure requires execution to detect.
The objective therefore supplies no preference among the partial scores whose ordering we validate, and it can reproduce distinctions made by imperfect tests.
We report discrimination separately on the Verified tasks excluded from strategy optimization (\cref{sec:populations,sec:disc}), which addresses task-level overlap but not the mismatch between the optimization objective and the score definition.
A rank-based objective grounded in independent quality judgments would align optimization more directly with the validated use of the score.

The rubric generator and judge may also have encountered Verified tasks or their reference solutions during training.
This is distinct from possible exposure of the systems that produced the submissions, which we treat as an observed population of candidate patches.
After excluding the two systems with the highest mean reference-patch line recall, mean per-task AUROC is 0.819 on the tasks that remain contested, compared with 0.826 for the full pool. %
Under token recall, mean per-task AUROC is 0.763 for resolved submissions written differently from the reference. %
Reference-like submissions receive higher scores, with the largest difference on the file-change axis (\cref{app:similarity}). %
These controls show that the reported discrimination is not limited to reference-like submissions, but they cannot determine whether the generator or judge recalled task-specific information or separate such recall from repository exploration.
Evaluation on tasks whose solutions postdate the evaluator models would provide a stronger control.

Existing public benchmarks do not provide all the properties needed for broader validation.
SWE-bench Verified supplies many comparable submissions per task, but covers only Python tasks whose reference patches are small on average.
SWE-bench Pro contains larger and more varied tasks \citep{deng2025swebenchpro}, but fewer comparable submissions are available.
Audits have identified problems with task specifications and test scope in both datasets \citep{openai2026verifiedretirement,openai2026proaudit}.
Both also rely on public tasks and solutions that may have been available during model training.
A broader validation dataset should vary task size, number of affected files, type of work, specification completeness, integration depth, and the number of plausible implementation strategies.
It should provide multiple candidate patches at different quality levels for each task, since within-task ordering cannot be evaluated from a reference patch alone.
Private or newly collected tasks would also reduce exposure to public solutions.

An independent dataset with multiple submissions and human per-task rankings could validate the ordering independently of test labels.
Annotators would rank which patch is closest to a finished solution, and agreement between annotators would provide context for interpreting agreement with the instrument.
Such rankings would compare failing patches directly, but would not validate score distances or establish that the rubric contains every requirement.
Combining them with independent requirement elicitation, criterion review, and verdict review would separately test ordering, item-sampling error, and rating error.

Broader task regimes may require changes to both rubric structure and the agent that generates it.
The current rubric is a flat set of criteria organized under four fixed axes and produced by one agent with repository access.
File-specific criteria help localize small repairs, but may overconstrain feature work that admits several valid architectures or file layouts.
Different types of work may be better served by different sets of axes.
Other possible refinements include criteria grouped under broader requirements, separate component and integration rubrics, and staged repository exploration.
A more elaborate rubric-generation harness could also incorporate mechanisms such as self-review of proposed criteria or searches for omissions.
These mechanisms could improve coverage on cross-cutting work.
They could also introduce additional criteria that unnecessarily constrain valid implementations.
Neither effect has been measured.

\section{Conclusion}
\label{sec:conclusion}

Tests are the most direct evidence that a patch works, but they cannot be assumed for every task or repository.
Repository-grounded rubrics score a patch without tests, an execution environment or a reference solution, and report which sampled requirements it appears to satisfy.
On SWE-bench Verified the score orders resolved above unresolved patches within a task, and its ranking of systems by their failing patches closely agrees with their ranking by resolve rate.
This validates the ordering against test outcomes, not absolute requirement satisfaction, and the case analyses show where judging from text and judging from tests each fall short.
Within those limits, the rubric score gives coding-agent evaluation a graded and diagnostic signal where tests are absent, incomplete or too costly to run.

\section*{Code and data availability}
\label{sec:availability}

The rubrics, the criterion verdicts behind the reported results, the rubric-generation and judge prompts, and a script that recomputes the validation results from these files are available at \url{https://github.com/faros-ai/repository-grounded-rubrics}.
The repository does not redistribute the benchmark tasks or submitted patches, but provides a script that downloads them from their public sources and verifies them against recorded hashes.

\section*{Acknowledgments}
\label{sec:acknowledgments}

We thank Juan Manuel Contreras, Kin Fai Kan and Ron Meldiner for their feedback and support.

\section*{Disclosure of LLM use}
\label{sec:llm-disclosure}

We used Claude (Anthropic) and Codex (OpenAI) to edit prose, write analysis and figure scripts, and prepare the LaTeX.
The authors checked every reported number against the underlying data and take full responsibility for the paper's content.
The use of language models as rubric generator and judge is part of the method described in \cref{sec:framework,sec:validation}.

\bibliographystyle{plainnat}
\bibliography{references}  %

\appendix
\crefalias{section}{appendix}
\crefalias{subsection}{subappendix}
\section{Rubric-generation strategy and optimization}
\label{app:optimization}

\subsection{Prompt composition}

The rubric generator receives a fixed system template, a fixed instance template, and a separately supplied strategy prompt.
The system template assigns the code-review role, requires a tool call in every turn, and appends the strategy text.
The instance template supplies the problem statement, repository path, size budget, output schema, and instructions for repository exploration and YAML validation.
It also prohibits access to version-control history and instructs the agent to create no files other than the rubric.
At run time, the two rendered templates form the initial system and user messages, respectively.

The templates were adapted from the agentic-rubric generator of \citet{raghavendra2026agenticrubrics}.
The system template retains their role and per-turn response requirements, and adds the reserved finishing steps and the slot into which the strategy is inserted.
The instance template retains their repository-exploration workflow, axes, and output format.
Our version additionally prohibits access to version-control history, accepts optional audit feedback, supplies a task-dependent size budget, uses task-general rather than bug-specific wording, removes the \texttt{underlying\_bug} metadata field, and specifies the submission command.
The fixed templates are reproduced in \cref{lst:rubric-system-template,lst:rubric-instance-template}, followed by the deployed strategy in \cref{lst:deployed-strategy}.
The listings use ASCII equivalents for typographic punctuation; their wording is otherwise unchanged.

\begin{lstlisting}[
	caption={The fixed system template for rubric generation. The strategy variable is replaced by the deployed strategy at run time.},
	label={lst:rubric-system-template}
]
You are an expert code reviewer that can understand issues and are well versed in the codebase. \
Your job is to write high-quality rubrics to grade the solution to a given issue.

IMPORTANT: In EVERY turn, you MUST ALWAYS include:
1. A summary of your thinking - explain what you're planning to do and why, and what tool \
you're going to use in (4-5 sentences max).
2. A tool call. You can only make one tool call per turn.

Leave at least 3 steps for writing, validating the YAML, and submitting.
{% if strategy %}

{{ strategy }}
{% endif %}
\end{lstlisting}

\begin{lstlisting}[
	caption={The fixed instance template for rubric generation. The variables are filled with the repository path, task description, optional audit feedback, and size budget.},
	label={lst:rubric-instance-template}
]
<uploaded_files>
{{ cwd }}
</uploaded_files>
I've uploaded a python code repository in the directory {{ cwd }}. Consider the following PR description:

<pr_description>
{{ task }}
</pr_description>

Can you help me write high quality rubrics to grade the solution to the task described in the <pr_description>?
This means you SHOULDN'T attempt to solve the task yourself, but rather understand the task, go through the codebase, and write rubrics only.

IMPORTANT: Do NOT inspect version-control history or metadata to find the reference solution - do not run `git log`, `git show`, `git diff`, `git blame`, or inspect branches, tags, stashes, or the reflog, and do not search for commits, PRs, or changelog entries that describe the fix. Write the rubric from the PR description and the current state of the code only.
{% if feedback %}

<reviewer_feedback>
A previous rubric you wrote for this exact task was audited and found defective. Address this feedback when writing the new rubric:

{{ feedback }}
</reviewer_feedback>
{% endif %}

<rubric_size_budget>
Aim for approximately {{ total_criteria }} rubric items in TOTAL across all four axes. This is a soft target -- distribute the items across the axes according to what the task genuinely requires, and do not pad or trim solely to hit the number.
</rubric_size_budget>

Follow these steps to write the rubrics:
1. As a first step, it might be a good idea to find and read code relevant to the <pr_description> by searching the codebase using search tools.
2. Then think of the approach to solve the task (functional requirements, non-functional requirements, etc.)
3. Also understand how the codebase is structured and how the code is organized, as well as coding style, etc.
4. Write a list of rubrics that can be used to grade the solution to the task described in the <pr_description>. Your rubrics should be along the axes described below.
5. Then finally, make a new file in the {{ cwd }} directory called `rubrics.yaml` with the rubrics you wrote. It should be a valid YAML file with the structure described later below
6. Also take a turn to ensure the yaml file is parseable by the yaml.safe_load function on the file itself. If it throws an error, fix the yaml file and take another turn to ensure it is parseable.
7. This should be the only file you create in the {{ cwd }} directory. DO NOT create/modify/delete any other files or directories.
8. Finally, when you are done and the rubrics.yaml file is valid, submit by running: echo COMPLETE_TASK_AND_SUBMIT_FINAL_OUTPUT

Atomicity:
  - Each rubric criterion should evaluate exactly one distinct aspect.
  - Avoid bundling multiple criteria into a single rubric. Most stacked criteria with the word "and" can be broken up into multiple pieces.

Self-containment & specificity (strict):
  - Do NOT write generic items; bind each item to exact paths/symbols/tokens seen in this instance.
  - Never rely on cross-item references; each item stands alone with its own identifiers and patterns.
  - The judge will only have access to the problem and patch and the current rubric under evaluation, so make sure the rubric can be evaluated without any other information.

Mutually Exclusive, Collectively Exhaustive (MECE):
  - The rubric set should be mutually exclusive and collectively exhaustive.

Style constraints (strict):
  - YAML only-no prose outside YAML.
  - Avoid backslash-heavy patterns; if you absolutely must include one, double any backslashes so the YAML stays valid.
  - Each rubric description starts with a third-person singular verb (e.g., Identifies, Implements, Validates, Confirms, Avoids, Cleans up, Plans).
  - Make descriptions concrete using tokens from PATCH/PR_DESCRIPTION.
  - Each rubric item includes: id (short), description (verb-first, instance-grounded), weight (int; 1=nice, 2=valuable, 3=must).
  - Avoid double-counting: do not re-score the same behavior under multiple items.

Pattern-writing guidelines (keep literal YAML-friendly text; no regex required):
  - Use plain path mentions like "diff --git a/path/to/file.py" or "+++ b/path/to/file.py".
  - Refer to symbols with straightforward phrases such as "def my_function(" instead of regex classes.
  - Describe value patterns in words (e.g., "string containing total") instead of complex expressions.
  - If you need to forbid something, just mention the exact string '@pytest.mark.skip', etc

Axes (execution-free):
  - file_change_rubrics (4 - 8): Scope, locality, and sufficiency of edits in PATCH (files/symbols/guards/regexes/flags). Penalize unrelated file churn; reward minimal, reversible diffs tied to the stated task.
  - spec_alignment_rubrics (3 - 6): Alignment of code to PR_DESCRIPTION. Use textual acceptance criteria (required types/conditions/error handling/API contracts) and ensure the patch reflects them.
  - integrity_rubrics (3 - 6): Hygiene and "no-cheating" safeguards-avoid test weakening (if tests appear in PATCH), mass renames, or dependency churn; preserve public API/semantics unless PR_DESCRIPTION requires otherwise.
  - runtime_rubrics (3 - 6): Natural-language criteria describing **intended runtime behavior** (NOT concrete tests), supported by execution-free textual evidence.
      - **Distinguishability:** Ensures the patch introduces or preserves signals that differentiate correct vs. incorrect behavior under plausible inputs (e.g., specific exception class, sentinel return, boundary guard).
      - **Regression safety:** Confirms backward-compatibility constraints (e.g., original API signatures/flags remain valid, deprecations gated via warnings).
      - **Determinism / flake resistance:** Avoids nondeterministic sources at runtime (unseeded randomness, wall-clock sleeps, network I/O) that would make tests flaky.
      - **Resource & timeout bounds:** Prevents pathological loops or heavy calls; respects existing timeouts/limits.
      - **Error-surface clarity:** Produces stable, specific messages/exception types that a test could assert against (not vague strings).
      - **Harness integrity:** Does not bypass or disable the project's runner/verifier hooks (e.g., keeps regression filters, CLI exit codes).

Return exactly this YAML structure:
metadata:
  task_summary: "<one-sentence summary grounded in PR_DESCRIPTION>"
axes:
  file_change_rubrics:
    - id: "FC1"
      description: "Identifies ..."
      weight: 3
    - id: "FC2"
      description: "Identifies ..."
      weight: 2
  spec_alignment_rubrics:
    - id: "SA1"
      description: "Recognizes ..."
      weight: 2
  integrity_rubrics:
    - id: "I1"
      description: "Confirms ..."
      weight: 2
  runtime_rubrics:
    - id: "R1"
      description: "Maintains ..."
      weight: 2
\end{lstlisting}

\subsection{Strategy selection}

The strategy prompt controls how the agent translates its repository inspection into criteria, including criterion scope, weighting, and the treatment of alternative implementations.
We used automated prompt search to refine a handwritten strategy rather than selecting every revision through manual iteration.
GEPA \citep{agrawal2025gepa} proposed revisions from natural-language feedback on evaluation failures and retained candidates with higher fitness.
The output schema, four axes, weight scale, tools, system template, and instance template remained fixed.
We chose GEPA because it was the prompt-optimization framework with which we had prior experience.
Other frameworks were available, and we did not compare them.

The optimization objective used candidate patches with known test verdicts.
For a task with candidate set $P$, let $g(p)\in\{0,1\}$ denote the test verdict for patch $p$ and let $S_R(p)$ be its score under a generated rubric $R$.
The per-patch objective was
\begin{equation}
	\label{eq:optimization-fitness}
	f(R,p)=\begin{cases}
		1 & \text{if } g(p)=1 \text{ and } S_R(p)=1, \text{ or } g(p)=0 \text{ and } S_R(p)<1,\\
		S_R(p) & \text{if } g(p)=1 \text{ and } S_R(p)<1,\\
		0 & \text{if } g(p)=0 \text{ and } S_R(p)=1.
	\end{cases}
\end{equation}
The fitness of a strategy on a task was the mean of $f(R,p)$ over $P$, and GEPA compared strategies by their mean fitness across tasks.
This objective rewards a rubric when test-passing patches receive the maximum score and test-failing patches receive less than the maximum.
Among failing patches below that boundary, it does not reward an ordering or a particular score.

The production run used a fixed corpus of 185 tasks and 809 candidate-patch records drawn from SWE-bench Verified and SWE-bench Pro.
The fixed split assigned 148 tasks to training and 37 to validation, without holding out repositories from the validation split.
Patches that could not be applied were excluded before optimization.
Following the model-family allocation of \citet{raghavendra2026agenticrubrics}, we used a Claude Sonnet model for repository-grounded rubric generation and a GPT model for judging, but selected newer versions available during development.
Rubrics and strategy revisions were generated with claude-sonnet-4-6, and rubric scores for the optimization objective were produced with gpt-5.1.
The validation study instead uses gpt-5.4 as its judge.
We did not systematically compare alternative generator or judge models.

The deployed strategy was selected from 16 candidates by its fitness on the optimization validation split.
Its mean fitness was 0.897, compared with 0.838 for the handwritten seed.
One validation task received zero fitness for every candidate because its repository could not be checked out.
These values describe selection under \cref{eq:optimization-fitness}; they are not a before-and-after comparison under the validation measures of \cref{sec:protocol}.
A later wording change replaced ``correctly fixes the described bug'' with ``correctly addresses the described task'' in the deployed strategy.

\subsection{Deployed strategy}

\begin{lstlisting}[
	caption={The deployed rubric-generation strategy. Typographic punctuation is rendered with ASCII equivalents.},
	label={lst:deployed-strategy}
]
Axis guidance:
- file_change_rubrics: scope, locality, and sufficiency of edits -- what the diff adds, removes, or modifies, and whether changes are confined to the appropriate locations.
- spec_alignment_rubrics: alignment of the code changes to the PR description's stated problem and acceptance criteria -- does the patch address what was broken?
- integrity_rubrics: "no-cheating" hygiene -- no test weakening, no mass renames, no dependency churn, public API preserved unless the PR description requires otherwise.
- runtime_rubrics: intended runtime behavior supported by execution-free textual evidence (no concrete tests).

Weight guidance:
- 3 = must-have, 2 = valuable, 1 = nice-to-have.
- When evaluating whether an item is a "must-have" (weight 3), ask: would a patch that omits this but correctly addresses the described task still be considered correct? If yes, do not use weight 3.
- Never assign weight >= 2 to items that require test additions or changelog/documentation updates. Use weight 1 at most for these ancillary artifacts.

Item shape:
- Atomicity: one distinct aspect per item; split "and"-joined criteria. Do not write compound multi-part items using "(a) ... (b) ... (c) ..." -- split into separate atomic items.
- MECE (mutually exclusive, collectively exhaustive) across the set: items do not overlap, and together cover the required behavior.
- Descriptions start with a third-person singular verb (e.g., Identifies, Implements, Validates, Confirms, Avoids) and use concrete tokens from the patch or the PR description.

## Critical: Verifiability from the diff alone

The judge sees ONLY the PR description and the patch diff -- not the repository, not the pre-patch source, not test output. Every item must be verifiable by reading the diff text.

**Do NOT write items that:**
- Assert runtime outcomes or numeric results (e.g., "produces values close to X" or "gives results in range Y") -- these cannot be read from a diff.
- Require executing the code to verify (e.g., "no longer raises", "dask lazy evaluation is preserved") unless there is a structural code pattern in the diff that directly implies it.
- Reference specific test file names, test method names, or repo-internal paths that a correct patch might legitimately differ from -- judge only the logic, not the location.
- Are so implementation-specific that alternative correct approaches would fail them (e.g., requiring one specific arithmetic expression when multiple equivalent forms exist, or requiring a specific function name, variable name, or encoding strategy when other approaches achieve the same result).
- Require knowing pre-patch state that is not visible in the diff itself (deletion hunks, PR description, or added lines must make both old and new state inferable).
- Pin a behavior to a specific call site or function when the PR description does not require it -- if a feature could be implemented in config loading, in the open function, or in a helper, do not require it to appear in any one of those places.

**Prefer items that:**
- Check structural properties visible in the diff: a branch or condition was added, a code path was split, a specific wrong value was removed, a new parameter or import was introduced.
- Describe added lines using concrete identifiers (class names, function names, import paths, string literals) that will literally appear in the diff.
- Describe removals via deletion hunks (lines prefixed `-`) anchored to an identifier or file name visible in the diff header.
- For behavioral items, point to a structural pattern in added code (e.g., "adds a conditional branch that checks dtype kind 'm'") rather than claiming a runtime outcome.
- For test additions, describe what the test logically exercises (e.g., "adds a test case exercising `polyval` with a `timedelta64` coordinate"), not the exact file path or parametrize ID, so equivalent placements pass.
- When describing WHERE logic must appear, allow any reasonable location -- use phrasing like "somewhere in the diff" or "in config loading or database opening code" rather than naming a specific function, unless the PR description explicitly requires a particular placement.

## Avoiding over-specification (the main cause of underrating correct patches)

When a PR description describes a behavior, there are often multiple valid implementation strategies. Write items that are satisfied by any equivalent approach. Use language like:
- "Adds code (anywhere in the diff) that constructs a connection string from the discrete fields..."
- "Adds a conditional check (in config loading or database setup code) that..."
- "Adds validation logic that returns an error containing a `db.`-prefixed field name when..."

Reserve location-specific items for cases where the PR description explicitly mandates a particular placement.

**Behavioral items to prefer over implementation-specific items:**
- Focus on what the fix achieves (the observable structural change and its purpose), not on one specific implementation strategy.
- "Removes the old X-only handling and replaces it with a mechanism that handles all Y cases" -- focuses on behavior, visible from deletion + addition hunks.
- "Adds code that decodes Z sequences" -- checks the goal without mandating one approach.

When multiple implementation strategies are valid, write items that pass any of them. Only anchor to a specific approach when the PR description explicitly mandates it.

## Calibrating weight against implementation likelihood

For items describing behaviors the PR explicitly requires, use weight 3 or 2. Only demote to weight 1 for nice-to-have details (documentation updates, minor phrasing consistency, test additions) that a minimal but correct implementation might omit. Do not make a weight-3 item out of a constraint that reasonable correct patches might implement differently (e.g., a specific error message format) -- keep those at weight 1-2 so underrating a correct patch has limited score impact.

## Calibrating strictness for test and changelog items

- For test additions: describe the coverage (what scenarios are tested) rather than the test function name or file location.
- For existing test updates: only write an item if the behavioral change in the production code logically requires test updates; describe the change in terms of observable test logic, not the test file or method name.
- Give test and changelog items weight 1 since they verify hygiene, not core correctness. A patch that correctly implements the required behavior is a correct solution even if it omits tests or a changelog entry.
\end{lstlisting}

\section{Choice of the rubric size rule}
\label{app:sizing}

We fixed the size budget of \cref{eq:size} after an exploratory comparison of three sizing rules.
The comparison used a convenience subset comprising all 44 \texttt{sphinx-doc/sphinx} tasks in SWE-bench Verified and predates the rubrics of \cref{sec:validation}. %
We did not select this repository as representative of the benchmark.
Under the first rule the agent chose the number of criteria itself, with no size directive in its prompt.
The second rule set the budget from the changed lines $L$ of the reference patch as $N \approx 8 + 6 \ln L$, without a cap.
The third is the capped linear rule of \cref{eq:size}.
The three rubric sets were generated under one configuration that differed only in the sizing rule, and the same 545 patches from 13 leaderboard systems, 183 of them resolved, were judged by gpt-5.4 under each rubric set. %
The reference patches of these tasks change a median of 7.5 lines and at most 107. %

\Cref{tab:sizing} reports the comparison.
Without a directive the agent wrote 10 to 16 criteria, and the count was weakly correlated with the changed lines (Pearson $r$ 0.12). %
Both formula rules tie the count to the changed lines, the logarithmic rule with roughly twice as many criteria as the linear one on this population. %
Per-task AUROC over the 32 contested tasks is 0.754, 0.754 and 0.751 under the three rules. %
The mean reference-patch scores are 0.825, 0.800 and 0.802. %
Judging each of the 44 reference patches three times gives a mean run-pair correlation of 0.825, 0.943 and 0.962, with criterion verdicts differing between runs for 5.4\%, 5.0\% and 3.7\% of reference-patch--criterion pairs. %

\begin{table}[h]
	\centering
	\small
	\setlength{\tabcolsep}{4.5pt}
	\caption{Three sizing rules on the 44 sphinx tasks. Criteria per rubric, Pearson correlation of the count with the reference patch's changed lines, mean reference-patch score, mean run-pair correlation over three judgments of each reference patch, and per-task AUROC over the 32 contested tasks. %
	}
	\label{tab:sizing}
	\begin{tabular}{lrrrrrr}
		\toprule
		Rule & Criteria (mean) & Range & $r$(criteria, lines) & Mean score & Retest $r$ & Per-task AUROC \\
		\midrule
		Agent-chosen                                  & 12.3 & 10--16 & 0.12 & 0.825 & 0.825 & 0.754 \\
		$8 + 6 \ln L$                                 & 22.8 &  9--39 & 0.82 & 0.800 & 0.943 & 0.754 \\
		$\min(60,\ \operatorname{round}(8 + L/12))$, \cref{eq:size}         & 10.4 &  9--19 & 0.98 & 0.802 & 0.962 & 0.751 \\
		\bottomrule
	\end{tabular}
\end{table}

The observed mean per-task AUROCs were similar, so the choice among the rules rested on the other columns.
We kept the linear rule because its count follows the task size most closely, with the fewest criteria, and its repeated judgments varied least.
Within the range tested here, the choice of sizing rule changed the number of criteria but produced little change in mean per-task discrimination.

The comparison is small.
It covers one repository, 44 tasks, 32 of them contested, and the tasks are small changes, so the linear rule produced between 9 and 19 criteria. %
The cap of 60 was never reached, and the comparison does not test whether the chosen slope remains appropriate for larger changes. %
The repeated judgments cover reference patches only, which sit at the top of the score range where most criterion verdicts are clear, so the differences in run-pair correlation are measured on a restricted range.
The rubrics of this comparison are not those of \cref{sec:validation}; the sphinx rubrics there were regenerated under the linear rule in September 2026 with the deployed strategy, and the comparison used a 13-system pool that predates the 15-system pool of \cref{tab:populations}, so its AUROC values are not comparable to those of \cref{tab:auc}. %
The generator model used for these earlier rubric sets was not recorded. %

\section{Validation population and rubric characteristics}
\label{app:population}

The fifteen leaderboard systems whose submissions we score were submitted between June 2024 and December 2025 and resolve between 25.2\% and 79.2\% of the tasks they attempted (\cref{tab:systems}). %
A system has no patch for a task when it generated none or published no evaluation log, which accounts for the 201 empty cells of the grid. %
Of those, 138 are missing generations, 61 missing logs and 2 empty patches. %
Three systems published a patch for all 500 tasks and the sparsest, \texttt{honeycomb}, for 439. %

\begin{table}[h]
	\centering
	\footnotesize
	\renewcommand{\arraystretch}{1.1}
	\caption{The fifteen SWE-bench Verified leaderboard systems of the validation population, ordered by resolve rate. \emph{Submitted} is the date of the system's leaderboard entry, \emph{patches} the tasks for which it published a patch, and \emph{no patch} the remainder of the 500. The resolve rate is over the patches a system published, not over 500, and the per-system statistics of \cref{sec:capability} use the same denominator. Counts are the harness verdicts each system published. Eight of the 7,299 patches exceed the judge's context window and have no rubric score, four from \texttt{tools\_claude-3-5-sonnet-updated} and four from \texttt{openhands\_4x\_scaled}.}
	\label{tab:systems}
	\begin{tabular}{@{}llrrrr@{}}
		\toprule
		System & Submitted & Patches & Resolved & Resolve rate & No patch \\
		\midrule
		\texttt{agentless\_MCTS-Refine-7B}          & 2025-06-27 & 461 & 116 & 25.2\% & 39 \\
		\texttt{sweagent\_gpt4o}                    & 2024-07-28 & 447 & 116 & 26.0\% & 53 \\
		\texttt{sweagent\_claude3.5sonnet}          & 2024-06-20 & 486 & 168 & 34.6\% & 14 \\
		\texttt{agentless-1.5\_gpt4o}               & 2024-10-28 & 496 & 194 & 39.1\% &  4 \\
		\texttt{swerl\_llama3\_70b}                 & 2025-02-26 & 499 & 206 & 41.3\% &  1 \\
		\texttt{gru}                                & 2024-08-24 & 497 & 226 & 45.5\% &  3 \\
		\texttt{honeycomb}                          & 2024-08-20 & 439 & 203 & 46.2\% & 61 \\
		\texttt{tools\_claude-3-5-sonnet-updated}   & 2024-10-22 & 483 & 245 & 50.7\% & 17 \\
		\texttt{agentless-1.5\_claude-3.5-sonnet-20241022} & 2024-12-02 & 497 & 254 & 51.1\% &  3 \\
		\texttt{devlo}                              & 2024-11-08 & 500 & 271 & 54.2\% &  0 \\
		\texttt{openhands\_4x\_scaled}              & 2025-02-03 & 498 & 304 & 61.0\% &  2 \\
		\texttt{sweagent\_claude-3-7-sonnet}        & 2025-02-25 & 497 & 312 & 62.8\% &  3 \\
		\texttt{augment\_agent\_v0}                 & 2025-03-16 & 500 & 327 & 65.4\% &  0 \\
		\texttt{zencoder\_ai}                       & 2025-04-30 & 499 & 350 & 70.1\% &  1 \\
		\texttt{sonar-foundation-agent\_claude-opus-4-5} & 2025-12-05 & 500 & 396 & 79.2\% & 0 \\
		\midrule
		all                                         &            & 7,299 & 3,688 & 50.5\% & 201 \\
		\bottomrule
	\end{tabular}
\end{table}

The leaderboard entries are system configurations rather than distinct base models.
Several share a harness or model, while others combine candidate generation and selection across multiple calls.
\Cref{tab:system-configurations} summarizes the submitter-provided metadata and descriptions in the official leaderboard records at a pinned repository revision \citep{swebenchexperiments2026}.
When those records do not identify a model or procedure, the table reports it as undisclosed rather than inferring it from the entry name.

\begin{table}[p]
	\centering
	\footnotesize
	\renewcommand{\arraystretch}{1.05}
	\newcommand{\configrowrule}{\addlinespace[1pt]\cmidrule[0.02em]{1-4}\addlinespace[1pt]}
	\newcommand{\submissionlink}[2]{\href{https://github.com/SWE-bench/experiments/tree/40f164d5b8f1d249bf95a6df8b74b577fd8e519d/evaluation/verified/#1}{#2}}
	\caption{Configurations of the fifteen leaderboard systems in the validation population, in the order used in \cref{tab:systems}. Configuration information is taken from the submitter-provided metadata and descriptions in the official leaderboard records. We did not independently reconstruct the systems. System names link to the corresponding submission records.}
	\label{tab:system-configurations}
	\begin{tabular}{@{}>{\raggedright\arraybackslash}p{3.3cm}>{\raggedright\arraybackslash}p{2.7cm}>{\raggedright\arraybackslash}p{3.5cm}>{\raggedright\arraybackslash}p{5.3cm}@{}}
		\toprule
		System & Harness & Model configuration & Submitter-reported procedure \\
		\midrule
		\submissionlink{20250627_agentless_MCTS-Refine-7B}{MCTS-Refine-7B} & MCTS-Refine-7B & MCTS-Refine-7B & Fine-tuned on issue-resolution reasoning traces produced through Monte Carlo tree search and refinement. The record does not describe the inference procedure. \\
		\configrowrule
		\submissionlink{20240728_sweagent_gpt4o}{SWE-agent + GPT-4o} & SWE-agent & GPT-4o (2024-05-13) & Uses SWE-agent's interface for repository browsing, editing, and code execution. \\
		\configrowrule
		\submissionlink{20240620_sweagent_claude3.5sonnet}{SWE-agent + Claude 3.5 Sonnet} & SWE-agent & Claude 3.5 Sonnet & Uses the same SWE-agent interface. The exact model snapshot is unclear because the current metadata names a snapshot released after the submission date. \\
		\configrowrule
		\submissionlink{20241028_agentless-1.5_gpt4o}{Agentless 1.5 + GPT-4o} & Agentless 1.5 & GPT-4o (2024-05-13) & Localizes edit locations, generates candidate repairs, and validates and reranks them. \\
		\configrowrule
		\submissionlink{20250226_swerl_llama3_70b}{SWE-RL} & Agentless Mini & Llama3-SWE-RL-70B & Uses an Agentless-derived scaffold with a model trained by reinforcement learning on software-evolution data. \\
		\configrowrule
		\submissionlink{20240824_gru}{Gru} & Gru & undisclosed & Uses a staged workflow for repository exploration, change planning, editing, and patch review. \\
		\configrowrule
		\submissionlink{20240820_honeycomb}{Honeycomb} & Honeycomb & undisclosed & The submission record does not describe the model or procedure. \\
		\configrowrule
		\submissionlink{20241022_tools_claude-3-5-sonnet-updated}{Tools + Claude 3.5 Sonnet} & Tools & Claude 3.5 Sonnet (2024-10-22) & Gives the model tools for file editing and shell commands. \\
		\configrowrule
		\submissionlink{20241202_agentless-1.5_claude-3.5-sonnet-20241022}{Agentless 1.5 + Claude 3.5 Sonnet} & Agentless 1.5 & Claude 3.5 Sonnet (2024-10-22) & Uses the Agentless localization, repair, and patch-validation pipeline. \\
		\configrowrule
		\submissionlink{20241108_devlo}{devlo} & devlo & Claude 3.5 Sonnet (2024-10-22) & The submission record identifies the model but does not describe the agent procedure. \\
		\configrowrule
		\submissionlink{20250203_openhands_4x_scaled}{OpenHands 4x Scaled} & OpenHands & Claude 3.5 Sonnet (2024-10-22) and o3-mini (2025-01-31) & Samples Claude 3.5 Sonnet three times and uses a fine-tuned verifier to rerank solutions. The record does not specify the role of o3-mini. \\ %
		\configrowrule
		\submissionlink{20250225_sweagent_claude-3-7-sonnet}{SWE-agent + Claude 3.7 Sonnet} & SWE-agent & Claude 3.7 Sonnet (2025-02-19) & The entry is labeled Review Heavy, but the submission record does not describe the review procedure. \\
		\configrowrule
		\submissionlink{20250316_augment_agent_v0}{Augment Agent v0} & Augment Agent v0 & Claude 3.7 Sonnet and o1 & Uses shell and file-editing tools, followed by an o1-based majority-voting step. \\
		\configrowrule
		\submissionlink{20250430_zencoder_ai}{Zencoder} & Zencoder & Claude 3.7 Sonnet (2025-02-19), o4-mini, and o3 & Runs four agent configurations using Claude and o4-mini, then uses o3 to select the submitted patch. \\ %
		\configrowrule
		\submissionlink{20251205_sonar-foundation-agent_claude-opus-4-5}{Sonar Foundation Agent} & Sonar Foundation Agent & Claude Opus 4.5 & Uses a LlamaIndex tool-calling agent with shell, text-editing, and symbol-search tools. \\
		\bottomrule
	\end{tabular}
\end{table}

\begin{figure}[h]
	\centering
	\includegraphics{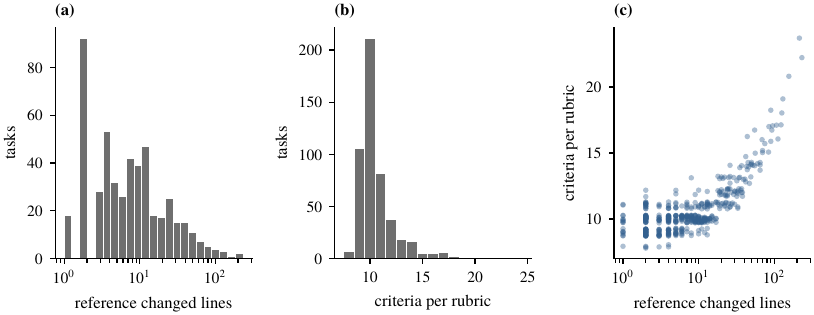} %
	\caption{The Verified population: (a) reference-patch changed lines, log scale; (b) criteria per rubric; (c) the two against each other, 500 tasks. %
	}
	\label{fig:profile}
\end{figure}

\Cref{fig:profile} shows the size of the reference change and the length of the rubric for the 500 SWE-bench Verified tasks of \cref{sec:populations}.
Two thirds of the reference patches change ten lines or fewer and 86\% touch a single file. %
The number of criteria follows the changed lines (Pearson $r$ 0.90 on the raw counts, Spearman 0.66) and only loosely the number of files ($r$ 0.37). %

\Cref{tab:auc} gives the per-task AUROC values plotted in \cref{fig:auc}, with the number of patches in contested tasks.

\begin{table}[h]
	\centering
	\small
	\caption{Per-task AUROC on SWE-bench Verified, resolved above unresolved within a task, the values plotted in \cref{fig:auc}. Contested tasks have at least one patch of each class and are the only ones the statistic is defined on. Intervals are 95\% bootstrap intervals over tasks.}
	\label{tab:auc}
	\begin{tabular}{lrrrl}
		\toprule
		Population & Tasks & Contested & Patches in contested & Mean per-task AUROC [95\% CI] \\
		\midrule
		Verified, all tasks               & 500 & 400 & 5,834 & 0.826 [0.805, 0.847] \\ %
		\quad not used in optimization    & 415 & 341 & 4,974 & 0.838 [0.817, 0.859] \\ %
		\quad used in optimization        &  85 &  59 &   860 & 0.760 [0.691, 0.822] \\ %
		\midrule
		django/django                     & 231 & 188 & 2,747 & 0.868 [0.844, 0.891] \\ %
		sympy/sympy                       &  75 &  59 &   872 & 0.799 [0.734, 0.863] \\ %
		sphinx-doc/sphinx                 &  44 &  35 &  503 & 0.822 [0.768, 0.869] \\ %
		matplotlib/matplotlib             &  34 &  27 &  368 & 0.765 [0.670, 0.854] \\ %
		scikit-learn/scikit-learn         &  32 &  27 &  402 & 0.767 [0.658, 0.863] \\ %
		pydata/xarray                     &  22 &  17 &  253 & 0.797 [0.678, 0.892] \\ %
		astropy/astropy                   &  22 &  15 &  222 & 0.789 [0.643, 0.912] \\ %
		pytest-dev/pytest                 &  19 &  17 &  250 & 0.875 [0.790, 0.936] \\ %
		pylint-dev/pylint                 &  10 &   5 &   73 & 0.793 [0.696, 0.891] \\ %
		psf/requests                      &   8 &   8 &  115 & 0.578 [0.409, 0.742] \\ %
		mwaskom/seaborn                   &   2 &   1 &   15 & 0.923 [0.923, 0.923] \\ %
		pallets/flask                     &   1 &   1 &   14 & 0.346 [0.346, 0.346] \\ %
		\bottomrule
	\end{tabular}
\end{table}

\section{Repeatability of rubric generation and judging}
\label{app:repeatability}

The score can vary because repeated judgments against a fixed rubric produce different criterion verdicts or because a new rubric contains a different sample of requirements.
The fixed-rubric study measures the first source, while repeated rubric generation measures their combined effect on scores and rankings.

\subsection{Fixed-rubric judge repeatability}

The judge-repeatability study used the 100-task manifest described in \cref{sec:repeatability}.
Each task contributed its lowest-, median- and highest-scoring non-reference submission under the original score, together with the reference patch.
We judged these 400 patches five times against a fixed rubric with the production judge configuration and with reasoning effort set to \emph{medium}. %

\Cref{tab:judge-repeatability} compares the two settings.
Pearson correlation and Kendall $\tau_b$ are means over the ten pairs of repeated runs.
Item disagreement is the fraction of patch--criterion pairs for which the five verdicts are not unanimous.
The ordering column counts tasks for which all four patches have the same ordering in every pair of runs.

\begin{table}[h]
	\centering
	\small
	\caption{Judge repeatability with fixed rubrics on 100 tasks and 400 patches. %
	}
	\label{tab:judge-repeatability}
	\begin{tabular}{lrrrrr}
		\toprule
		Reasoning effort & Pearson $r$ & Mean SD & Item disagreement & Kendall $\tau_b$ & Exact ordering \\
		\midrule
		None   & 0.985 & 0.020 & 4.8\% & 0.960 & 63/100 \\
		Medium & 0.964 & 0.030 & 6.9\% & 0.923 & 57/100 \\
		\bottomrule
	\end{tabular}
\end{table}

The production setting has lower observed variability on each pooled measure in the table.
This comparison supports the choice of no reasoning effort for this judge configuration, but does not compare the accuracy of the two settings.

\subsection{Rubric-generation repeatability}

We compared three independently generated rubrics for each task in the same manifest.
All three used claude-sonnet-4-6, the deployed strategy and prompts, a 50-step and \$15-per-task budget, and isolated repository checkouts. %
We label the resulting generations R1--R3.

Aggregate rubric length is stable across the three generations.
The mean criterion counts are 10.42, 10.47 and 10.44, with a median of 10 in each generation, even though the prompt gives the agent a soft target rather than a fixed criterion count. %
The mean share of rubric weight assigned to each axis differs by at most 0.005. %
The particular criteria nevertheless change enough to affect scores and rankings.
\Cref{tab:rubric-repeatability} reports pairwise agreement in two comparisons.
The first three agreement columns use all 1,451 labeled submissions, each judged once with each rubric generation.
The final two use four patches per task, each judged five times with each generation. %

\begin{table}[h]
	\centering
	\small
	\caption{Agreement between rubric generations. Score correlation, absolute score difference and submission-ordering $\tau_b$ use all 1,451 labeled submissions. Four-patch $\tau_b$ compares the five-judgment mean score under each rubric; exact ordering is the number of tasks with identical four-patch average-rank vectors. %
	}
	\label{tab:rubric-repeatability}
	\begin{tabular}{lrrrrr}
		\toprule
		Generation pair & Score $r$ & Mean $|\Delta|$ & Submission $\tau_b$ & Four-patch $\tau_b$ & Exact ordering \\
		\midrule
		R1--R2 & 0.866 & 0.114 & 0.677 & 0.747 & 33/100 \\
		R1--R3 & 0.871 & 0.112 & 0.677 & 0.732 & 37/100 \\
		R2--R3 & 0.873 & 0.112 & 0.669 & 0.750 & 34/100 \\
		\bottomrule
	\end{tabular}
\end{table}

Aggregate discrimination is more stable than individual scores and rankings.
Mean per-task AUROC over the 77 contested tasks is 0.842, 0.842 and 0.818 across the three generations. %
The bootstrap intervals for the pairwise differences include zero, so the study does not detect a clear difference in mean AUROC.
Task-level AUROC correlations of 0.58 to 0.74 show that this aggregate similarity does not imply the same result on each task. %

We also compare the two observed sources of variability on the same 400 patches.
Within-rubric variability is the standard deviation of five repeated judgments for each patch and rubric.
Between-rubric variability is the standard deviation of the three five-judgment mean scores for the same patch.
After averaging each quantity over the four patches of each task, the mean task-level standard deviation is 0.0200 within rubrics and 0.0814 between rubric generations, a ratio of 4.06. %

\section{Reference similarity and discrimination}
\label{app:similarity}

The rubric generator model of \cref{sec:populations} was trained after every SWE-bench Verified task was published, so its criteria could describe a remembered solution rather than the requirements of the problem.
If they did, the rubric would recognize code shaped like the reference patch and little else, and its agreement with the tests in \cref{sec:disc} would rest on resolved patches that happen to share that shape.
This appendix tests that reading.

We measure how much each resolved patch resembles the reference patch of its task, split the resolved patches into those written like the reference and those written differently, and compute the per-task AUROC of \cref{sec:protocol} twice, each group against every unresolved patch of the same task.
A rubric that rewarded only reference-likeness would separate the second group from the failures no better than chance.
A rubric that scores requirements would separate both.

Resemblance is measured on the code itself, from two perspectives.
Both patches are read as diffs and only the added lines of Python source files are kept, with test files and build artifacts dropped.
Only added lines are compared, because the removed lines of a patch are fixed by the code that already exists at the place it edits, so two patches that edit the same region remove the same lines whatever they add.
The added lines are the part of a patch its author wrote, and they are what a resemblance measure should read.
\emph{Line recall} reads the text.
Each added line has its whitespace removed so that indentation and spacing do not break a match, and lines of three characters or fewer are then dropped, which removes blank lines and lone brackets that any patch contains; 13\% of the reference patches' added source lines go this way, nearly two thirds of them empty. %
The recall of a candidate is the share of the reference patch's remaining distinct added lines that appear, as exact strings, anywhere among the candidate's added lines.
A renamed variable breaks the match, so this perspective is strict about form (\cref{lst:line}).
\emph{Token recall} reads the shape.
Each added line is split into tokens by the Pygments Python lexer, which tolerates the incomplete lines a diff contains, and normalized in the manner token-based clone detection normalizes identifiers and literals before comparison \citep{roy2009clones}.
Names become \texttt{ID}, numbers \texttt{NUM} and string literals \texttt{STR}, so that renaming and changed constants do not separate two copies, while keywords, builtins, operators and punctuation keep their text so that control flow does.
We depart from that scheme in one place: attribute names, the token after a dot, also keep their text, so that a call to a different method on the same object counts as different code.
The unit of comparison is the $n$-gram of normalized tokens, with $n=3$ and windows taken within one line, since single tokens are shared by any two Python patches and four-token windows break on every small edit.
The recall of a candidate is the share of the reference patch's trigrams, counted with multiplicity, that the candidate also contains.
A copy of the reference patch with renamed variables scores near 1 under token recall and near 0 under line recall, which is the difference between the two perspectives; a fix that uses different operations scores near 0 under both.
Under either measure a resolved patch is \emph{written differently} when its recall is below 0.5 and \emph{reference-like} otherwise.

\begin{lstlisting}[float=tbp,language=Python,caption={Line recall. \texttt{parse\_diff} returns the added lines of each file in a unified diff and \texttt{is\_source\_file} applies the file filter of the text.},label={lst:line}]
import re

def normalize_line(line: str) -> str:
    """Remove every whitespace character from an added line."""
    return re.sub(r"\s+", "", line)

def added_lines(patch: str) -> set[str]:
    """Set of normalized added lines in the source files of a diff, dropping very short lines."""
    lines = set()
    for path, added in parse_diff(patch).items():
        if is_source_file(path):
            lines.update(s for s in map(normalize_line, added) if len(s) > 3)
    return lines

def line_recall(gold: set[str], candidate: set[str]) -> float:
    """Share of the gold's added lines present in the candidate."""
    return len(gold & candidate) / len(gold)
\end{lstlisting}

\begin{lstlisting}[float=tbp,language=Python,caption={Token trigram recall and the per-task AUROC. \texttt{parse\_diff} returns the added lines of each file in a unified diff and \texttt{is\_source\_file} applies the file filter of the text.},label={lst:tokens}]
import builtins
import keyword
from collections import Counter
from pygments.lexers import PythonLexer
from pygments.token import Comment, Name, Number, String, Text, Whitespace

KEPT_NAMES = set(keyword.kwlist) | set(dir(builtins)) | {"self", "cls"}
LEXER = PythonLexer()

def normalize_tokens(line: str) -> list[str]:
    """Tokens of one line with names blanked to ID, numbers to NUM and string literals to STR."""
    out = []
    previous = ""
    for kind, text in LEXER.get_tokens(line):
        if kind in Text or kind in Whitespace or kind in Comment:
            continue
        if kind in Name.Decorator:
            out.append(text)
        elif kind in Name:
            out.append(text if text in KEPT_NAMES or previous == "." else "ID")
        elif kind in Number:
            out.append("NUM")
        elif kind in String:
            if not out or out[-1] != "STR":
                out.append("STR")
        else:
            out.append(text)
        previous = text
    return out

def trigrams(patch: str) -> Counter:
    """Multiset of normalized token trigrams over the added source lines of a diff."""
    counts: Counter = Counter()
    for path, added in parse_diff(patch).items():
        if not is_source_file(path):
            continue
        for line in added:
            tokens = normalize_tokens(line)
            counts.update(zip(tokens, tokens[1:], tokens[2:]))
    return counts

def token_recall(gold: Counter, candidate: Counter) -> float:
    """Share of the gold's trigrams, counted with multiplicity, present in the candidate."""
    return sum((gold & candidate).values()) / sum(gold.values())

def auroc(positives, negatives) -> float:
    """Normalized Mann-Whitney U: probability a positive outscores a negative, ties count half."""
    wins = sum((p > n) + 0.5 * (p == n) for p in positives for n in negatives)
    return wins / (len(positives) * len(negatives))
\end{lstlisting}

\begin{figure}[htbp]
	\centering
	\includegraphics{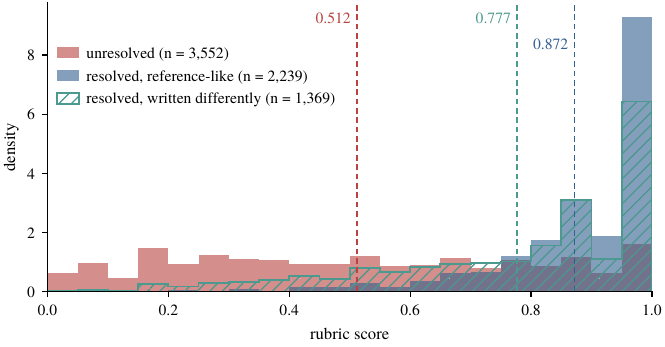} %
	\caption{Score distributions of unresolved patches, resolved patches written differently from the reference, and reference-like resolved patches under token trigram recall, 7,160 scored patches over 491 SWE-bench Verified tasks; dashed lines mark the group means (0.512, 0.777, 0.872). %
	}
	\label{fig:anchor}
\end{figure}

Under line recall 62\% of the resolved patches on Verified are written differently from the reference, and under token recall 38\%; the two agree on 73\% of the resolved patches, and nearly all of the disagreement is renamed or lightly rewritten copies that line recall calls different and token recall calls reference-like. %
On the differently written patches the mean per-task AUROC against the unresolved patches of the same task is 0.800 [0.775, 0.823] over 355 tasks under line recall and 0.763 [0.733, 0.791] over 276 tasks under token recall; on the reference-like patches it is 0.884 [0.857, 0.909] and 0.872 [0.849, 0.893]. %
On the tasks that have both kinds of resolved patch the pairs are 0.789 against 0.911 over 198 tasks and 0.759 against 0.885 over 198. %
The result does not depend on the cut: at a threshold of 0.25 the AUROC on the differently written patches is 0.782 under line recall and 0.708 under token recall, and at 0.75 it is 0.814 and 0.789. %
\Cref{tab:similarity} collects the values and \cref{fig:anchor} shows the three score distributions behind them.

\begin{table}[t]
	\centering
	\small
	\caption{Per-task AUROC of resolved patches against the unresolved patches of the same task, by whether the resolved patch resembles the reference. Written differently: recall below 0.5. Interval: 95\% bootstrap over tasks. Last column: the two AUROCs restricted to tasks that have resolved patches of both kinds, so that they rest on the same tasks.}
	\label{tab:similarity}
	\begin{tabular}{@{}lrllr@{}}
		\toprule
		Measure & Written differently & AUROC, written differently & AUROC, reference-like & Both kinds, same tasks \\
		\midrule
		Line recall          & 62\% & 0.800 [0.775, 0.823] (355) & 0.884 [0.857, 0.909] (240) & 0.789 vs 0.911 (198) \\ %
		Token trigram recall & 38\% & 0.763 [0.733, 0.791] (276) & 0.872 [0.849, 0.893] (316) & 0.759 vs 0.885 (198) \\ %
		\bottomrule
	\end{tabular}
\end{table}

Resembling the reference does earn a premium.
Reference-like resolved patches score higher than differently written ones, 0.872 against 0.777 on average, and the gap is widest on the file-change axis, 0.793 against 0.680. %
This is the over-specification of \cref{sec:framework} made visible: some criteria describe where and how the reference patch made its change rather than what any correct change must do, and patches that took the same route collect them.
It is a bonus for looking like the reference, worth about 0.1 of AUROC, and not the carrier of the discrimination in \cref{sec:disc}. %

We also measured reference similarity by candidate system, averaging line recall over each system's resolved patches.
The mean is 0.51 for the system built on Claude Opus 4.5 and 0.78 for the 7B model fine-tuned on SWE-bench-style data, compared with 0.26 to 0.38 for the other thirteen systems. %
Using only the patches from those two systems gives a mean per-task AUROC of 0.919 [0.889, 0.947]. %
Excluding their patches and recomputing per-task AUROC on the tasks that remain contested gives 0.819 [0.797, 0.839] on 356 tasks, compared with 0.826 [0.805, 0.847] on 400 tasks for the full pool. %
Shared preference for the reference's implementation choices is one possible explanation for the higher agreement in the two-system subset, but this comparison does not isolate its cause and the subsets contain different contested tasks.
The exclusion check establishes that discrimination persists without the systems whose resolved patches most resemble the reference.

The test bounds the effect that any recall of the reference solution could have on discrimination; it does not establish whether recall occurred.
Every Verified task was published before the generator model was trained, so a criterion that names the reference patch's file could come from memory or from reading the repository, and this appendix cannot tell the two apart.
Text similarity is also a proxy for a patch being written differently, not for it being a different solution, since two correct fixes of a seven-line bug will share tokens by necessity. %
\Cref{sec:limitations} states the residual and the population on which it could be resolved.

\section{Disagreement case studies}
\label{app:cases}

This appendix presents the evidence behind the case studies in \cref{sec:disagreement}.
Each case connects the task description and relevant candidate changes to the benchmark outcome and the rubric's assessment.
The examples illustrate tests enforcing an unstated requirement, tests missing a regression, an execution-blocking defect missed by the evaluation, and inconsistent judgments of equivalent changes.
They illustrate these mechanisms without establishing their prevalence.

We reproduce the relevant diff excerpts, test assertions or harness errors, and criterion verdicts.
Criterion texts are abridged for presentation, and comment lines are omitted from diff excerpts.

\subsection{\texttt{django-16667}: tests enforce an unstated requirement}

The issue reports that \texttt{SelectDateWidget.value\_from\_datadict} raises \texttt{OverflowError} when a date component is larger than \texttt{sys.maxsize}.
The candidate, from \texttt{sonar-foundation-agent\_claude-opus-4-5}, widens the existing handler.

\begin{lstlisting}
-            except ValueError:
+            except (ValueError, OverflowError):
                 return "%s-%s-%s" % (y or 0, m or 0, d or 0)
\end{lstlisting}

The reference adds a second handler with a fixed return value.

\begin{lstlisting}
+            except OverflowError:
+                return "0-0-0"
\end{lstlisting}

The test patch asserts the reference's value in \texttt{test\_value\_from\_datadict}, \texttt{((str(sys.maxsize + 1), "12", "1"), "0-0-0")}, and adds \texttt{f.clean("0-0-0")} to \texttt{test\_datefield\_1}; the candidate returns \texttt{"9223372036854775808-12-1"} for the same input.
Thirteen of the fifteen systems add \texttt{OverflowError} to the existing handler. %
Nine restrict their edits to \texttt{widgets.py} and score 1.0; the other four also edit additional files and score 0.909. %
No system resolves the task. %

\begin{table}[h]
	\centering
	\small
	\caption{Rubric for \texttt{django-16667}, verdicts for the candidate (score 1.0) and the reference (0.773).} %
	\label{tab:case16667}
	\begin{tabular}{lp{8.1cm}rcc}
		\toprule
		Id & Criterion (abridged) & Weight & Candidate & Reference \\
		\midrule
		FC1 & Modifies only \texttt{django/forms/widgets.py} among production files & 2 & 1 & 1 \\
		FC2 & Adds \texttt{OverflowError}, alone or with \texttt{ValueError}, to the except clause & 3 & 1 & 1 \\
		FC3 & Leaves the pseudo-ISO fallback return reachable for both error types & 2 & 1 & 0 \\
		SA1 & Targets the try/except around \texttt{datetime.date(int(y), int(m), int(d))} & 3 & 1 & 1 \\
		SA2 & An oversized year no longer propagates \texttt{OverflowError} & 3 & 1 & 1 \\
		I1 & Invalid but non-overflowing components still return the fallback string & 3 & 1 & 1 \\
		I2 & Does not weaken the widget's existing tests & 1 & 1 & 1 \\
		R1 & Returns the pseudo-ISO fallback string instead of raising on overflow & 3 & 1 & 0 \\
		R2 & Valid dates still return a formatted date string & 2 & 1 & 1 \\
		\bottomrule
	\end{tabular}
\end{table}

\subsection{\texttt{django-14122}: tests miss a regression}

The issue asks that \texttt{Meta.ordering} fields no longer be included in the \texttt{GROUP BY} clause.
The reference, in \texttt{django/db/models/sql/compiler.py}, guards the loop that adds order-by columns to the grouping.

\begin{lstlisting}
-        for expr, (sql, params, is_ref) in order_by:
-            if not is_ref:
-                expressions.extend(expr.get_group_by_cols())
+        if not self._meta_ordering:
+            for expr, (sql, params, is_ref) in order_by:
+                if not is_ref:
+                    expressions.extend(expr.get_group_by_cols())
\end{lstlisting}

The candidate, from \texttt{gru}, replaces the order-by contribution in \texttt{compiler.py} with a filter and adds a similar filter to the existing grouping expressions in \texttt{query.py}.

\begin{lstlisting}
-                expressions.extend(expr.get_group_by_cols())
+                ordering_fields = {field for field in self.query.get_meta().ordering}
+                expressions = [expr for expr in expressions if expr not in ordering_fields]
\end{lstlisting}

For the test model, the field-name string and unresolved ordering expressions in \texttt{Meta.ordering} do not match the column expressions being grouped, so the filters remove nothing.
Removing the \texttt{extend} call suppresses the grouping contribution of order-by expressions, including explicit ordering; columns already contributed by selected fields remain grouped.
The sole \texttt{FAIL\_TO\_PASS} test, \texttt{test\_default\_ordering\_does\_not\_affect\_group\_by}, aggregates over a model whose only ordering is \texttt{Meta.ordering}, and passes.
The designated regression tests pass as well.
Fourteen systems submitted a patch, and nine of them resolve the task. %

\begin{table}[h]
	\centering
	\small
	\caption{Rubric for \texttt{django-14122}, verdicts for the candidate (score 0.083) and the reference (0.958).} %
	\label{tab:case14122}
	\begin{tabular}{lp{8.1cm}rcc}
		\toprule
		Id & Criterion (abridged) & Weight & Candidate & Reference \\
		\midrule
		FC1 & Edits limited to the order-by-to-\texttt{GROUP BY} logic in \texttt{compiler.py} & 3 & 0 & 1 \\
		FC2 & No files changed outside \texttt{compiler.py} and related tests & 2 & 0 & 1 \\
		FC3 & Adds a test for aggregation on a model with \texttt{Meta.ordering} & 1 & 0 & 0 \\
		SA1 & Guards \texttt{get\_group\_by} against expressions that originate from \texttt{Meta.ordering} & 3 & 0 & 1 \\
		SA2 & Uses \texttt{self.\_meta\_ordering} to identify those expressions & 3 & 0 & 1 \\
		SA3 & Keeps the existing suppression of \texttt{Meta.ordering} from \texttt{ORDER BY} & 2 & 0 & 1 \\
		I1 & Explicit \texttt{order\_by()} calls still contribute to \texttt{GROUP BY} & 3 & 0 & 1 \\
		I2 & No existing aggregation tests removed or weakened & 2 & 1 & 1 \\
		R1 & Aggregation on a model with \texttt{Meta.ordering} excludes the ordering fields from \texttt{GROUP BY} & 3 & 0 & 1 \\
		R2 & Order-by entries that come only from \texttt{Meta.ordering} contribute no \texttt{GROUP BY} columns & 2 & 0 & 1 \\
		\bottomrule
	\end{tabular}
\end{table}

SA3, R1 and R2 describe behavior the candidate's diff preserves or produces, and a verdict of 1 on each would be defensible, while SA1 and SA2 describe the reference's mechanism rather than a requirement.
I1 nevertheless identifies a regression that the designated tests miss; the aggregate score should not be interpreted as a measure of that regression alone.

\subsection{\texttt{astropy-8707}: a high-scoring patch that cannot import}
\label{app:case-astropy8707}

The issue asks for \texttt{Header.fromstring} and \texttt{Card.fromstring} to accept bytes as well as strings.
The candidate from \texttt{gru} adds ASCII decoding to those methods, but also edits \texttt{\_BaseHDU.fromstring} in \texttt{astropy/io/fits/hdu/base.py} and \texttt{HDUList.fromstring} in \texttt{astropy/io/fits/hdu/hdulist.py}.
The change to \texttt{base.py} includes a stray fragment of a method signature.

\begin{lstlisting}
+        if isinstance(data, bytes):
+            data = data.decode('ascii')
+                   **kwargs):
\end{lstlisting}

The harness fails while importing that module during test collection.
Its traceback identifies the malformed line.

\begin{lstlisting}
E       **kwargs):
E   IndentationError: unexpected indent
\end{lstlisting}

The recorded rubric score is 0.943. %
The judge marks every integrity and runtime criterion satisfied.
The integrity criteria ask whether the public signatures of \texttt{Header.fromstring} and \texttt{Card.fromstring} are preserved and whether existing tests are left intact.
The runtime criteria concern decoding bytes before string operations in those methods; none explicitly checks syntactic validity of the patched files.
These verdicts recognize the requested local changes without accounting for the import failure.
The malformed line is visible in the supplied diff and could be detected by a syntax check; the framework performs no such independent check.

\subsection{\texttt{django-10914}: the judge misreads the patch}

The issue asks for a default of \texttt{0o644} for \texttt{FILE\_UPLOAD\_PERMISSIONS}.
The reference changes one line of \texttt{django/conf/global\_settings.py}.

\begin{lstlisting}
-FILE_UPLOAD_PERMISSIONS = None
+FILE_UPLOAD_PERMISSIONS = 0o644
\end{lstlisting}

Fourteen systems submitted a patch, and twelve make the same settings replacement, eleven with the same replacement line and one with a trailing comment. %
\texttt{sweagent\_gpt4o} replaces \texttt{DEBUG = False} near the beginning of \texttt{global\_settings.py} with the new assignment, leaving a later assignment to \texttt{None} that overrides it; \texttt{swerl\_llama3\_70b} edits \texttt{django/core/\allowbreak files/storage.py} instead.
These patches are excluded from the comparison of equivalent settings changes.
Several of the twelve add documentation or a test file beside the line.
The candidate, from \texttt{augment\_agent\_v0}, adds a paragraph and a note to \texttt{docs/topics/http/file-uploads.txt}.
Four criteria, SA1, I1, R1 and R2, are decided by the settings line alone, since neither the reference nor the candidate touches \texttt{FileSystemStorage}.
\Cref{tab:case10914} lists the rubric with the verdicts for the candidate and the reference, and \cref{tab:case10914systems} the four verdicts for every patch that makes the same effective settings replacement.

\begin{table}[h]
	\centering
	\small
	\caption{Rubric for \texttt{django-10914}, verdicts for the candidate (score 0.143) and the reference (0.762). FC3, SA2 and I2 ask for documentation and test changes that neither patch contains.} %
	\label{tab:case10914}
	\begin{tabular}{lp{8.1cm}rcc}
		\toprule
		Id & Criterion (abridged) & Weight & Candidate & Reference \\
		\midrule
		FC1 & Changes \texttt{FILE\_UPLOAD\_PERMISSIONS} from \texttt{None} to \texttt{0o644} in \texttt{global\_settings.py} & 3 & 1 & 1 \\
		FC2 & Confines the fix to \texttt{global\_settings.py} without modifying unrelated files & 2 & 0 & 1 \\
		FC3 & Updates the default in \texttt{docs/ref/settings.txt} & 2 & 0 & 0 \\
		SA1 & With the default, \texttt{FileSystemStorage} applies \texttt{0o644} on both upload paths & 3 & 0 & 1 \\
		SA2 & Revises the \texttt{settings.txt} text describing the old inconsistency & 1 & 0 & 0 \\
		I1 & An explicitly configured permission is still applied in \texttt{\_save} & 3 & 0 & 1 \\
		I2 & Updates the test for the default case to \texttt{0o644} & 2 & 0 & 0 \\
		R1 & Both upload paths end at \texttt{0o644} under the new default & 3 & 0 & 1 \\
		R2 & Setting \texttt{None} explicitly still opts out of \texttt{chmod} & 2 & 0 & 1 \\
		\bottomrule
	\end{tabular}
\end{table}

\begin{table}[!htb]
	\centering
	\small
	\caption{The four line-decided criteria of \texttt{django-10914} across the twelve patches that make the reference's effective settings change, with what else each patch changes and its score.}
	\label{tab:case10914systems}
	\begin{tabular}{@{}lp{5.0cm}ccccr@{}}
		\toprule
		System & Other files changed & SA1 & I1 & R1 & R2 & Score \\
		\midrule
		\texttt{gru} & none & 1 & 1 & 1 & 1 & 0.762 \\
		\texttt{agentless-1.5\_gpt4o} & none & 1 & 1 & 1 & 1 & 0.762 \\
		\texttt{devlo} & none & 1 & 1 & 1 & 1 & 0.762 \\
		\texttt{agentless-1.5\_claude-3.5-sonnet-20241022} & none & 0 & 0 & 0 & 0 & 0.238 \\
		\texttt{sonar-foundation-agent\_claude-opus-4-5} & three documentation files & 1 & 1 & 1 & 1 & 0.762 \\
		\texttt{sweagent\_claude-3-7-sonnet} & two documentation files & 1 & 1 & 1 & 1 & 0.810 \\
		\texttt{zencoder\_ai} & \texttt{docs/ref/settings.txt} & 0 & 0 & 0 & 0 & 0.238 \\
		\texttt{augment\_agent\_v0} (candidate) & \texttt{docs/topics/http/\allowbreak file-uploads.txt} & 0 & 0 & 0 & 0 & 0.143 \\
		\texttt{sweagent\_claude3.5sonnet} & \texttt{docs/topics/http/\allowbreak file-uploads.txt} & 0 & 0 & 0 & 0 & 0.143 \\
		\texttt{honeycomb} & a stray \texttt{patch.diff}; line carries a comment & 0 & 0 & 0 & 0 & 0.143 \\
		\texttt{tools\_claude-3-5-sonnet-updated} & a new test file & 0 & 0 & 0 & 0 & 0.143 \\
		\texttt{openhands\_4x\_scaled} & a new test file & 0 & 0 & 0 & 0 & 0.143 \\
		\bottomrule
	\end{tabular}
\end{table}

\section{Cost of the instrument}
\label{app:cost}

We measured the cost of rubric generation on the 100 SWE-bench Verified tasks of the repeatability study (\cref{sec:repeatability}) by generating one new rubric per task. %
We used the validation study's generator claude-sonnet-4-6, the strategy prompt of \cref{app:optimization}, and the agent budget of 50 steps and \$15 per task (\cref{app:repeatability}). %
Anthropic prompt caching was enabled for this run.
These rubrics are a new sample from the same generator and are not the rubrics judged in \cref{sec:validation}, and the figures below come from this single run.
We computed dollar amounts from the provider's per-call token counts and list prices rather than from invoices. %
The prices were \$3 per million input tokens, \$0.30 per million read from the cache, \$3.75 per million written to it, and \$15 per million output tokens. %
All 100 generations completed, and one reached the step ceiling and still produced a rubric. %

A rubric cost \$0.25 at the median and \$0.28 on average, with a 90th percentile of \$0.45 and a maximum of \$0.61 (\cref{tab:cost}). %
The agent took 27 steps at the median. %
Re-pricing the same calls without caching, with every prompt token at the list input rate, gives a median cost of \$1.16 per rubric, 4.6 times the cached figure. %
This calculation re-prices the measured tokens rather than measuring an uncached run. %
An agent without caching may take a different number of steps. %
We also compared cached and uncached generation on the same 44 sphinx tasks under the same configuration. %
Across the two runs, total cost on those tasks was \$41.22 without caching and \$13.77 with caching, a factor of 3.0. %
The median cost per rubric was \$0.90 and \$0.29, respectively. %
Because the runs followed independent agent trajectories, the ratio includes run-to-run variation in exploration. %
Pearson $r$ between generation cost and the number of changed lines $L$ in the reference patch was 0.31. %

\begin{table}[h]
	\centering
	\small
	\caption{Rubric generation cost by repository on the 100 tasks of \cref{sec:repeatability}. Cost per rubric in US dollars with prompt caching and median agent steps. %
	}
	\label{tab:cost}
	\begin{tabular}{lrrrr}
		\toprule
		Repository & Tasks & Median cost & Mean cost & Median steps \\
		\midrule
		sphinx        & 44  & 0.29 & 0.31 & 30.5 \\
		django        & 32  & 0.26 & 0.26 & 26.5 \\
		sympy         & 11  & 0.25 & 0.25 & 21.0 \\
		matplotlib    &  4  & 0.26 & 0.27 & 25.5 \\
		scikit-learn  &  3  & 0.13 & 0.13 & 14.0 \\
		astropy       &  2  & 0.20 & 0.20 & 20.5 \\
		xarray        &  2  & 0.26 & 0.26 & 26.0 \\
		pytest        &  1  & 0.16 & 0.16 & 16.0 \\
		requests      &  1  & 0.15 & 0.15 & 14.0 \\
		\midrule
		All           & 100 & 0.25 & 0.28 & 27.0 \\
		\bottomrule
	\end{tabular}
\end{table}

The production judge, gpt-5.4, makes one model call per candidate.
The judging runs took place between August 20 and September 8, 2026, and the pipeline priced their calls at \$2.50 per million input tokens and \$15 per million output tokens. %
Ninety-two of the 8,481 judge calls in the validation runs, which also cover the reference patches and patches from systems outside the fifteen-system pool, cost less than these rates give, consistent with part of the prompt priced as cached input at \$0.25 per million. %
Across the 1,451 submission judgments on the same 100 tasks, one judgment cost \$0.0052 at the median. %
A task had 15 labeled submissions at the median, and judging all of them cost \$0.084 at the median, about a third of generating the task's rubric. %
Spearman $\rho$ between judging cost and the character count of the diff was 0.68, compared with 0.43 between cost and the number of criteria. %
Within each quartile of patch length, rubrics of twelve or more criteria cost between \$0.0012 and \$0.0017 more per judgment at the median than rubrics of nine or fewer. %
The measured instrument cost for $N$ candidates on a task is one rubric generation plus $N$ judgments.
Once the rubrics of a task set exist, scoring patches from a second harness, a new model version, or a best-of-$N$ sweep adds only the judging cost.

The pipeline records the cost of the turns whose model output it retained.
If a model turn contains a tool call that fails to parse, the provider bills it but the pipeline does not record it.
No such omission occurred in this run. %
Recorded steps equaled billed calls on every task, so the totals above include every billed call. %

\end{document}